\documentclass[pra,aps,twocolumn,superscriptaddress,nofootinbib,nopacs]{revtex4} %---APS---SI---arxiv
\usepackage{amsmath}  \usepackage{amssymb}  \usepackage{amsfonts}  \usepackage{bm}  \usepackage{bbm}   \usepackage{braket}  \usepackage{color}  \usepackage{comment}  \usepackage{dcolumn}  \usepackage{enumerate}  \usepackage{epsfig}  \usepackage{gensymb}  \usepackage{graphicx}  \usepackage{indentfirst}  \usepackage{lmodern}  \usepackage{mathrsfs}  \usepackage{mathtools}  \usepackage{psfrag}  \usepackage{pst-all}  \usepackage{soul}  \usepackage{units}  \usepackage{xcolor}
\usepackage{float} %---APS---SI---arxiv
\usepackage[colorlinks,linkcolor=blue,citecolor=blue,urlcolor=blue,hyperindex,driverfallback=dvipdfm]{hyperref}  \usepackage[T1]{fontenc} %---APS---ACS---SI---arxiv

\def\ii{{\rm i}}  \def\ee{{\rm e}}

                  \def\gb{{\bf g}}                          \def\Rb{{\bf R}}        \def\vb{{\bf v}} %--- bold vectors
                
  \def\red{\color{red}}    
\def\wL{\omega_L}  \def\Om{\Omega}  \def\db{{\bf d}}
\def\tw{\tilde\omega}  \def\trho{\rho}  \def\rrho{\tilde\rho}

\begin{document} %---APS---SI---arxiv
%\renewcommand{\thefigure}{S\arabic{figure}} %---SI
%\renewcommand{\theequation}{S\arabic{equation}} %---SI
%\renewcommand{\thetable}{S\arabic{table}} %---SI
%\renewcommand{\thesection}{S\arabic{section}} %---SI
%\renewcommand{\thepage}{S\arabic{page}} %---SI
%\def\bibsection{\section*{\refname}} %---SI---arxiv

% =========================================================
% --- title, affiliations, abstract -----------------------
% =========================================================
\title{Atomic Floquet Physics Revealed by Free Electrons}

% --- APS affiliations ------------------------------------
\author{Eduardo~Arqu\'e~L\'opez}
\affiliation{ICFO-Institut de Ciencies Fotoniques, The Barcelona Institute of Science and Technology, 08860 Castelldefels (Barcelona), Spain}
\author{Valerio~Di~Giulio}
\affiliation{ICFO-Institut de Ciencies Fotoniques, The Barcelona Institute of Science and Technology, 08860 Castelldefels (Barcelona), Spain}
\author{F.~Javier~Garc\'{\i}a~de~Abajo}
\email{javier.garciadeabajo@nanophotonics.es}
\affiliation{ICFO-Institut de Ciencies Fotoniques, The Barcelona Institute of Science and Technology, 08860 Castelldefels (Barcelona), Spain}
\affiliation{ICREA-Instituci\'o Catalana de Recerca i Estudis Avan\c{c}ats, Passeig Llu\'{\i}s Companys 23, 08010 Barcelona, Spain}

% --- document format -------------------------------------
%\begin{document} %---ACS
%\ociscodes{XXX} %(300.6530) Ultrafast spectroscopy; (270.1670) Coherent optical effects.} %---OSA

% --- abstract --------------------------------------------
\begin{abstract}
We theoretically investigate the ability of free electrons to yield information on the nonlinear Floquet dynamics of atomic systems subject to intense external illumination. By applying a quantum-mechanical formalism to describe the atom-electron interaction under the presence of a monochromatic classical light field, we observe multiple energy features that reveal a large departure from non-illumination conditions in two- and three-level illuminated atoms, including the emergence of energy features associated with direct electron-photon exchanges, as well as a varied set of Floquet resonances that strongly depend on the light intensity and frequency. Our results unveil a wealth of effects associated with the interaction between free electrons and optically driven electronic systems.
\end{abstract}

% --- document format -------------------------------------
%\setboolean{displaycopyright}{true} %---OSA
%\begin{document} %---OSA
\maketitle %---APS---OSA---SI---arxiv
\date{\today} %---APS---arxiv
%\tableofcontents %---APS---SI---arxiv optional
%\setcounter{equation}{0} %---OSA
%\setkeys{acs}{maxauthors=0} %---ACS to avoid "et al." in references
%\noindent \textbf{Keywords:} nonreciprocal media, electron microscopy, cathodoluminescence, terahertz spectroscopy, EELS %---ACS

% =========================================================
\section{Introduction}

Thanks to recent advances in electron microscope instrumentation \cite{BDK02,KLD14,KDH19} that enable electron energy-loss spectroscopy (EELS) to be performed with few-meV spectral resolution \cite{KLD14,KDH19,LTH17,LB18,HNY18,HHP19,HKR19,paper342,SSB19,HRK20,YLG21,paper369}, free electron beams (e-beams) allow us to map the atomic structure of materials with sub-{\AA}ngstrom spatial resolution \cite{NP98}, while additionally granting us access into the atomic-scale properties of electronic and phononic excitations at frequencies extending down to the mid-infrared. When the electrons are extracted from a photocathode, simultaneous exposure of the specimen to synchronized laser and electron pulses trigger ultrafast interactions that can be used to shape \cite{GLW06,BPK08,BFZ09,ARM20} and compress \cite{BZ07,SCI08,PRY17,MB18_2,KES17,KSH18,MB18,SMY19,RTN20} the electron wave function along directions parallel \cite{BFZ09,paper151,PLZ10,PLQ15,FES15,EFS16} and transverse \cite{paper311,paper312,paper332,FYS20} with respect to the e-beam. Such interactions have also been investigated as a way to map the temporal evolution of optical excitations in the so-called photon-induced near-field electron microscopy (PINEM) \cite{BFZ09,paper151,PLZ10,PZ12,KGK14,PLQ15,FES15,paper282,EFS16,KSE16,RB16,VFZ16,paper272,PRY17,KML17,FBR17,paper306,paper311,paper312,paper325,paper332,K19,PZG19,paper339,RML20,DNS20,KLS20,WDS20,RK20,MVG20,paper360,VMC20,KDS21,HRF21} technique, which has been applied, for example, to image the femtosecond dynamics of plasmon \cite{PLQ15,paper282} and phonon-polariton \cite{KDS21} optical fields. These studies have focused on optical excitations of bosonic nature, in which the materials respond linearly to the fields associated with both light and electrons, while the activation of nonlinear sample response driven by intense laser illumination can affect the electron spectra, potentially enabling the measurement of nonlinear response functions with nanoscale spatial resolution \cite{paper347}.

Among nonlinear systems, two-level atoms constitute a paramount example in which the absorption of one photon blocks subsequent absorption events. In addition, atoms subject to intense light irradiation develop a nonlinear Floquet dynamics characterized by time-evolving Stark shifts \cite{S1965}. Floquet physics has also been explored in a wide range of systems that range from electron bands in solids \cite{LRG11,SML15} and cold atoms in optical lattices \cite{JMD14} to photonic structures \cite{RZP13} and thermal distributions of excited nuclei \cite{BJA21}. The interaction of shaped e-beams with atomic systems has recently attracted attention because of its quantum nature \cite{GY20,ZSF21}, and although the excitation probability is known to be independent of the electron wave function \cite{paper371}, the phase of the transition amplitude can depend on the electron probability density \cite{paper374,paper373}. Electron beams also constitute a potentially interesting tool to investigate Floquet dynamics in atomic-scale electronic systems. In a related context, the scattering of low-energy electrons by illuminated atoms has been experimentally demonstrated to produce electron energy combs associated with multiple photon exchanges similar to PINEM \cite{WHS1983}. However, we expect additional features in the scattered electron spectra related to transitions among Floquet states, which can be incommensurate with the employed photon energy. Ultrafast electron microscopy offers an ideal platform to explore these phenomena.

In this work, we theoretically demonstrate that the energy spectra of electrons interacting with illuminated atoms display not only an energy comb of gain-loss EELS sidebands similar to PINEM (i.e., peaks associated with a net number of exchanged photons), but also features at energies that are incommensurate with the photon energy, emanating from transitions between the optically induced Floquet states. Multiple avoided crossings are observed in the dispersion diagram of these energy features as a function of illumination frequency and energy loss, while the evolution with varying light intensity reveals the effect of dynamical Stark shifts. The number of spectral peaks is substantially increased in three-level systems and exhibit a dramatic dependence on which of the electron- and light-driven transitions are allowed between atomic energy levels. The present study supports the use of free electrons to gain insight into the nonlinear Floquet dynamics of illuminated atomic systems, with predictions that could be tested in gas-phase targets, as well as in exciton-supporting materials, such as two-dimensional crystals in which EELS studies have already been performed \cite{TLM15}.

% =========================================================
\section{Theoretical formalism}

We consider a sample hosting $N$ electronic states $|a\rangle$ of energies $\hbar\varepsilon_a$ subject to monochromatic light irradiation and described by the Hamiltonian
\begin{align}
\hat{\mathcal{H}}_s(t)=\sum_a \hbar\varepsilon_a|a\rangle\langle a|+\cos(\wL t)\sum_{aa'}\hbar\Om_{aa'}|a\rangle\langle a'|,
\nonumber
\end{align}
where $\Om_{aa'}=\Om_{a'a}$ are optical Rabi frequencies (proportional to the light intensity) and $\wL$ is the light frequency. The system admits $N$ independent Floquet eigenstates \cite{S1965} $\ee^{-\ii\tw_jt}|F_j(t)\rangle$, where $\tw_j$ are characteristic frequencies and
\begin{align}
|F_j(t)\rangle=\sum_{al} \ee^{-\ii l\wL t} f_{jal}|a\rangle
\label{Fjmain}
\end{align}
form a complete ($\sum_j|F_j(t)\rangle\langle F_j(t)|=\sum_a|a\rangle\langle a|$) and orthonormal ($\langle F_j(t)|F_{j'}(t)\rangle=\delta_{jj'}$) set of states with the same periodicity as the optical drive. We include inelastic decay of the excited states through a Lindbladian incorporating small incoherent rates compared to both $\wL$ and the excitation energies. Then, the dynamics of the illuminated system prior to electron interaction follows a density matrix
\begin{align}
\hat\rho_s(t)=\sum_{jj'}\trho_{jj'}(t)\,|F_j(t)\rangle\langle F_{j'}(t)|,
\label{rhosmain}
\end{align}
where the time-dependent coefficients $\trho_{jj'}(t)=\sum_l\trho_{jj'l}\ee^{-\ii l\wL t}$ have the same periodicity as the applied light field (see Appendix\ \ref{App1}).

Electron-sample interaction is described through a Hamiltonian $\hat{\mathcal{H}}_{\rm e-s}=\sum_{aa'qq'}\db_{aa'}\cdot\gb_{q'-q}|aq\rangle\langle a'q'|$, where the sum runs over atomic transition dipoles $\db_{aa'}$ and electron momentum states $|q\rangle$ and $|q'\rangle$, coupled through the vectors $\gb_{q'-q}$ \cite{paper221}. We assume well focused and collimated electrons, so we only need to consider $q$ along the e-beam direction.  In addition, we adopt the following approximations: (i) sample inelastic transitions are sufficiently slow as to be neglected during the interaction time; (ii) the electron-sample coupling is weak enough to be treated at the lowest-order level of perturbation theory; (iii) the incident electron energy width is small compared to the photon energy; and (iv) the electron velocity vector remains nearly constant during the interaction time (nonrecoil approximation). Assumption (i) depends on the choice of sample, but can be easily satisfied for sub-picosecond electron pulses and sample inelastic processes dominated by radiative decay, while approximations (ii)-(iv) are generally applicable for resonances in the visible regime and e-beams in electron microscopes \cite{paper371}.

Under these conditions, the transmitted electron spectrum is characterized by inelastic peaks emerging at energies
\begin{align}
\hbar\omega_{jj'l}=\hbar(\tw_j-\tw_{j'}+l\wL),
\label{wjjlmain}
\end{align}
which correspond to differences between the Floquet frequencies $\tw_j$ supplemented by multiples of the incident photon frequency $\wL$. More precisely, after interaction with the sample, the EELS probability reduces to (see detailed derivation in Appendix\ \ref{App2})
\begin{align}
\Gamma_{\rm EELS}(\omega)\approx P_{\rm ZLP}\Gamma_{\rm ZLP}(\omega)+{\sum_{jj'l}}' P_{jj'l}\,\Gamma_{\rm ZLP}(\omega-\omega_{jj'l}) \nonumber
\end{align}
as a function of energy loss $\hbar\omega$, where $\Gamma_{\rm ZLP}(\omega)$ denotes the zero-loss-peak (ZLP) distribution of the incident electrons, $P_{\rm ZLP}=1-{\sum_{jj'l}}' P_{jj'l}$ is the fraction of electrons remaining in the ZLP, and the primed sums are restricted to either $j\neq j'$ or $l\neq0$. While we formulate a general theory in Appendix\ \ref{App2}, simpler results are obtained if the transition dipoles $\db_{aa'}$ are oriented perpendicularly to the e-beam and the ratio $R_e/v$ of the electron-sample separation to the electron velocity is small compared with $\omega_{jj'l}$. Then, the coupling vectors $\gb_{q'-q}$ turn out to be independent of $q'-q$, $R_e$, and $v$, so the EELS peak probabilities reduce to (see Appendix\ \ref{App2})
\begin{align}
\frac{P_{jj'l}}{P^0}=I_{jj'l}\sum_{j''l'} I_{jj''l'}\;
{\rm Re}\{\trho_{j''j',l'-l}\}, \label{Pjjlmain}
\end{align}
where the matrix elements
\begin{align}
I_{jj'l}=\sum_{aa'l'}f_{jal'}\,f_{j'a',l'-l}\,d_{aa'}/d_{10} \nonumber
\end{align}
are computed from the coefficients of the Floquet states and the steady-state density matrix in Eqs.\ (\ref{Fjmain}) and (\ref{rhosmain}), and we normalize $P_{jj'l}$ to the probability $P^0$ of the only loss peak obtained for a non-illuminated system in the two- and three-level atoms considered below.

% =========================================================
\section{Results and discussion}

% Figure 1 ------------------------------------------------
\begin{figure*}
\centering{\includegraphics[width=0.65\textwidth]{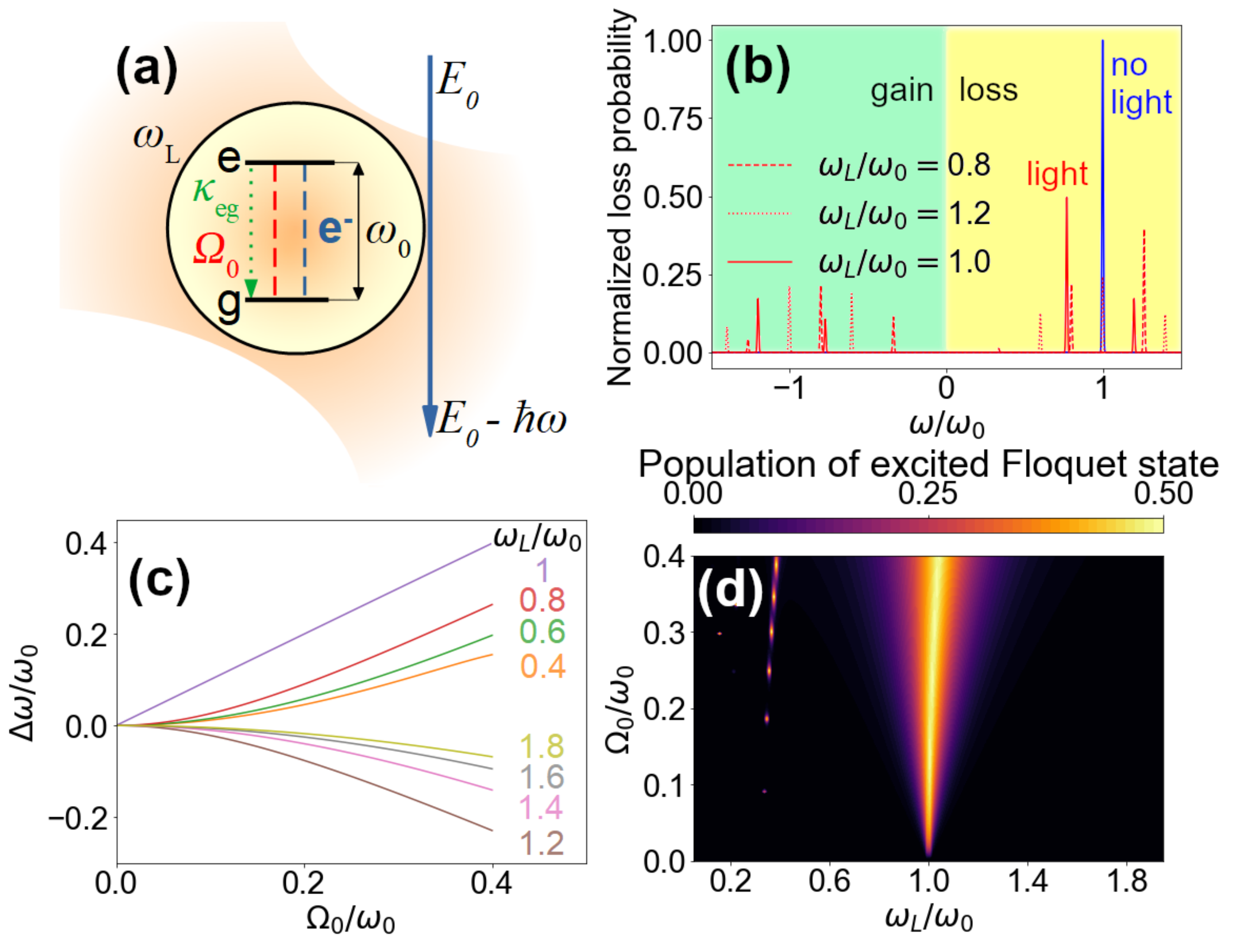}}
\caption{{\bf Free-electron interaction with an illuminated two-level atom.} {\bf (a)} Scheme of the system under consideration, consisting of an atom (excitation frequency $\omega_0=\varepsilon_e-\varepsilon_g$, decay rate $\kappa_{eg}\ll\omega_0$) subject to continuous-wave illumination (frequency $\wL$, Rabi frequency $\Om_0=\Om_{eg}$) and probed by a free electron. {\bf (b)} EELS probability in the absence (blue curve) and presence (red curves) of external illumination for different values of the light frequency $\wL$ and fixed strength $\Om_0=0.4\,\omega_0$, showing both loss ($\omega>0$) and gain ($\omega<0$) features as a function of energy loss $\hbar\omega$. {\bf (c)} Dynamical Stark shift of the Floquet transition frequency relative to the atomic frequency (i.e., $\Delta\omega=|\tw_1-\tw_0|-\omega_0$) as a function of $\Om_0$ for different light frequencies. {\bf (d)} Average population of the upper Floquet state $\rho_{110}$ as a function of $\wL$ and $\Om_0$. We normalize $\Omega_0$, $\omega$, and $\wL$ to $\omega_0$ in these plots.}
\label{Fig1}
\end{figure*}

For a two-level system [Fig.\ \ref{Fig1}(a), atomic states $a=g,e$] in the absence of illumination, there is just a single excitation frequency $\omega_0\equiv\varepsilon_e-\varepsilon_g$ that shows up as a loss peak in the EELS spectrum [Fig.\ \ref{Fig1}(b), blue curve]. Under strong illumination with Rabi frequency $\Om_0\equiv\Omega_{eg}=0.4\,\omega_0$, this feature is dramatically reduced and accompanied by peaks emerging at spectral positions that depend on the light frequency $\wL$ [Fig.\ \ref{Fig1}(b), red curves, calculated from Eq.\ (\ref{Pjjlmain})]. As anticipated, examining the $\Om_0$ dependence of the differences between the two Floquet frequencies [Fig.\ \ref{Fig1}(c)], we find monotonically increasing dynamical Stark shifts whose sign and magnitude depend on $\wL$. Also, the population of the upper Floquet state (i.e., the time-averaged value $\langle\rho_{11}(t)\rangle=\rho_{110}$ for $j=1$ [see Eq.\ (\ref{rhosmain})], which coincides with $|e\rangle$ in the $\Om_0=0$ limit) increases with $\Om_0$ and eventually reaches a maximum peak level of $\sim1/2$ (indicative of a Rabi oscillation regime) at $\wL$ blue shifted with respect to $\omega_0$ [Fig.\ \ref{Fig1}(d)]. Weaker population maxima are also observed, associated with harmonic excitation emerging for $\wL$ near the odd fractions of $\omega_0$ [see $\wL\sim\omega_0/3$ feature in Fig.\ \ref{Fig1}(d)].

% Figure 2 ------------------------------------------------
\begin{figure*}
\centering{\includegraphics[width=0.8\textwidth]{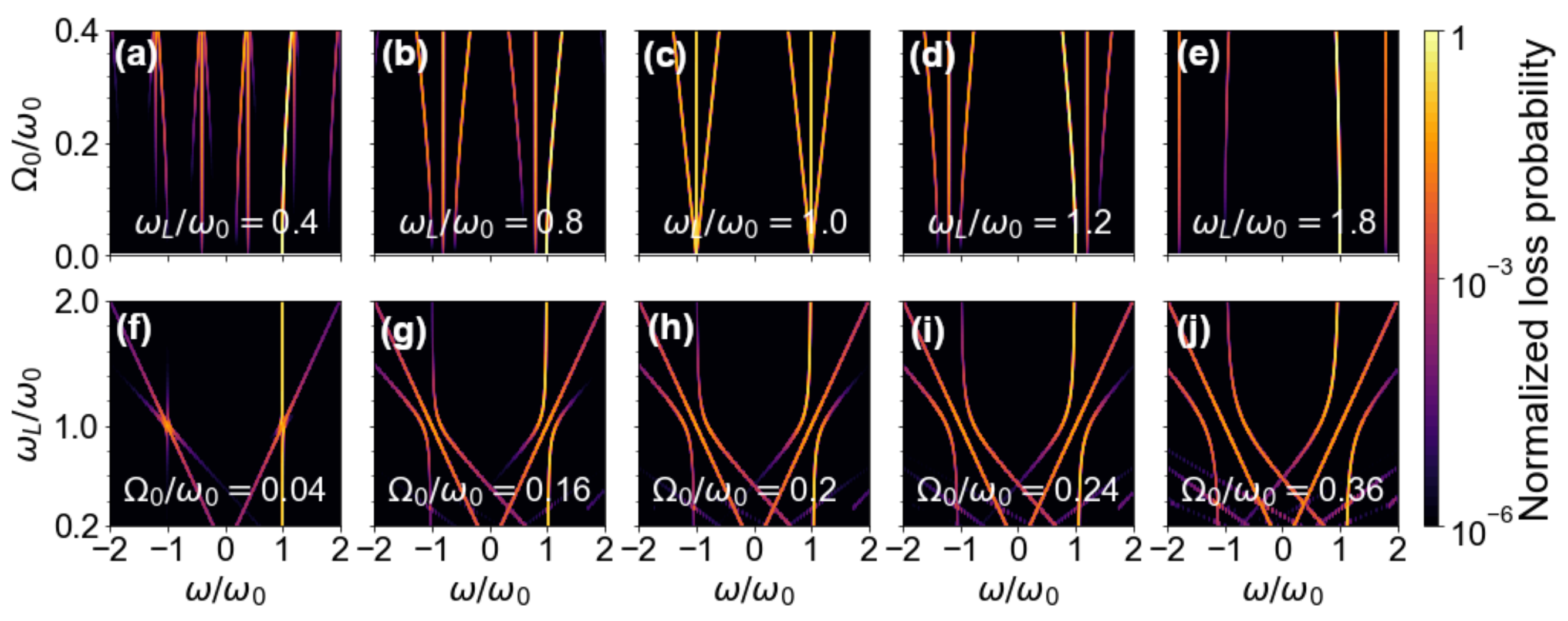}}
\caption{{\bf Probing Floquet resonances through EELS.} {\bf (a-e)} Evolution of the EELS probability as a function of energy loss (horizontal axis) and light-atom coupling strength $\Om_0$ (vertical axis) for the system depicted in Fig.\ \ref{Fig1}(a) with different illumination frequencies $\wL$ (see labels). {\bf (f-j)} Evolution of the EELS probability with $\wL$ (vertical axis) for different values of $\Om_0$ (see labels). We normalize $\Om_0$, $\omega$, and $\wL$ to $\omega_0$. The loss probability is broadened with a Gaussian of $0.01\,\omega_0$ full width at half maximum and normalized to the probability for the non-illuminated atom ($\Om_0=0$) at $\omega=\omega_0$.}
\label{Fig2}
\end{figure*}

The atomic excitations under consideration satisfy a simple selection rule: from the energies described by Eq.\ (\ref{wjjlmain}), Floquet features emerge at frequencies $\omega\pm(\tw_1-\tw_0)+l\wL$ with even $l$, while PINEM-like sidebands are observed at $\omega=l\wL$ with odd $l$ (i.e., $j=j'$ requires an even number of atom-light scattering events). When examining spectra for fixed $\wL$ and varying $\Om_0$ [Fig.\ \ref{Fig2}(a)-(e)], PINEM features are found to emerge as vertical lines, while Floquet transitions undergo Stark shifts with increasing $\Om_0$ in agreement with the behavior described in Fig.\ \ref{Fig1}(c). When varying the light frequency for fixed intensity [Fig.\ \ref{Fig2}(f)-(j)], PINEM sidebands form intense diagonal lines at $\omega=\pm\wL$ (i.e., with $l=\pm1$) accompanied by weaker higher-$l$ sidebands, while Floquet resonances evolve nonlinearly as a function of $\wL$, exhibiting a set of avoided and non-avoided crossings. For completeness, we reproduce Fig.\ \ref{Fig2} in the SI including $jj'l$ labels to identify each of the EELS peaks.

% Figure 3 ------------------------------------------------
\begin{figure}
\centering{\includegraphics[width=0.45\textwidth]{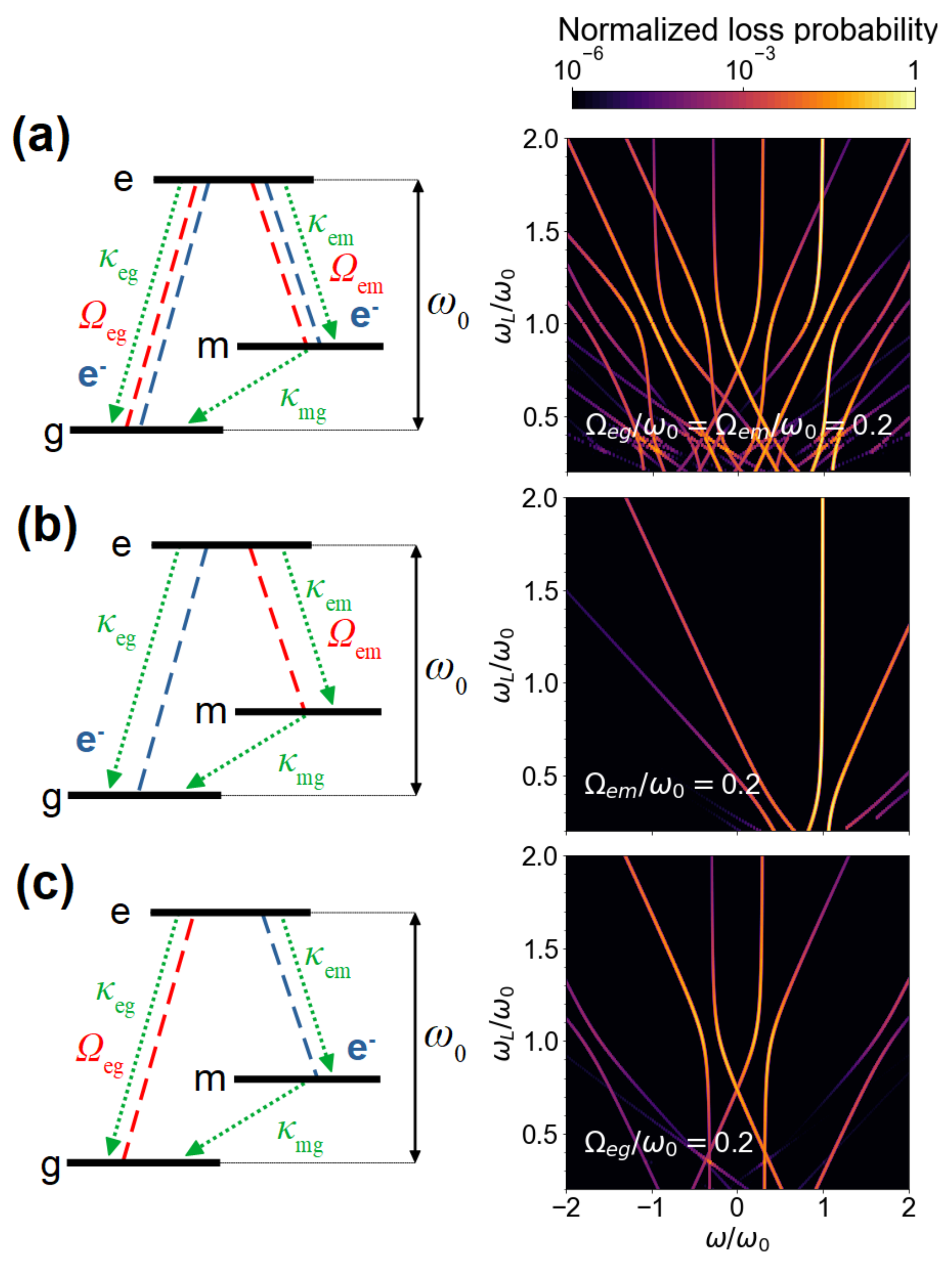}}
\caption{{\bf Free-lectron probing of Floquet resonances in tree-level atoms.} We consider the $\Lambda$-type atoms sketched in the left insets with $\omega_0=\varepsilon_e-\varepsilon_g$, $\omega_1=\varepsilon_e-\varepsilon_m=0.3\,\omega_0$, and three different configurations depending on which transitions ($e\leftrightarrow{g}$ and/or $e\leftrightarrow{m}$) are coupled to the light (red dashed lines) and the electrons (blue dashed lines), including inelastic decay as indicated by the green arrows. The right color plots show the evolution of the corresponding EELS probabilities with energy loss (horizontal axes) and light frequency $\wL$ (vertical axes). For simplicity, we set all decay rates and coupling rates to common values $\kappa_{eg}=\kappa_{em}=\kappa_{mg}\ll\omega_0$ and $\Om_{eg}=\Om_{em}$ (see labels), respectively. We follow the same normalization as in Fig.\ \ref{Fig2}.}
\label{Fig3}
\end{figure}

More complicated atomic systems also display interesting Floquet dynamics, particularly when the light and the electrons couple to different transitions. This situation is explored in Fig.\ \ref{Fig3} for a $\Lambda$-type atom that hosts an additional intermediate-energy state $|m\rangle$ and in which $e\leftrightarrow{g}$ and $e\leftrightarrow{m}$ transitions are selectively enabled for electrons and light, as indicated in the left sketches by blue and red dashed lines, respectively. For concreteness, we take $\omega_0=\varepsilon_e-\varepsilon_g$, $\omega_1=\varepsilon_e-\varepsilon_m=0.3\,\omega_0$, $\kappa_{eg}=\kappa_{em}=\kappa_{mg}\ll\omega_0$, and $d_{eg}=d_{em}$. If the upper excited state $|e\rangle$ can be reached from the two low-lying states through both electron and light excitation [Fig.\ \ref{Fig3}(a)], the spectral EELS probability, plotted as a function of light frequency $\wL$ (vertical axis) in the color plot, displays similar spectral peaks as those observed in the two-level atom in Fig.\ \ref{Fig2}(h) for the same Rabi frequency, but now supplemented by additional excitations involving the presence of a third state. When the ground state can be excited by electrons alone, while light only couples the two excited states [Fig.\ \ref{Fig3}(b)], the number of EELS features is substantially reduced, and in particular, no PINEM sidebands are preserved. Obviously, in this particular scenario the system lies in the ground state before interacting with the electron, and $|g\rangle$ is in fact one of the Floquet states ($j=0$), but surprisingly, the electron undergoes energy gains [features at $\omega=\tw_j-\tw_0+l\wL<0$ in Fig.\ \ref{Fig3}(b)] with finite probability $P_{j0l}/P^0=I^2_{j0l}$ by transitioning to excited states ($j>0$) through the Floquet ladder ($l<0$). The effect of the external illumination is also revealed by an avoided energy crossing affecting the $g\rightarrow{e}$ transition when the photon frequency matches $\wL\sim\varepsilon_e-\varepsilon_m$. In the opposite scenario [$g\leftrightarrow{e}$ coupling through light and $m\leftrightarrow{e}$ through the electron, Fig.\ \ref{Fig3}(c)], PINEM sidebands are also absent, but we find signatures of gain and loss near the $\pm(\varepsilon_e-\varepsilon_m)$ transition frequencies, accompanied by strongly Stark-splitted peaks. Obviously, the dynamics of these systems is highly nonlinear, so the features observed in Fig.\ \ref{Fig3}(b)-(c) when removing some of the couplings considered in Fig.\ \ref{Fig3}(a) are not just a subset of the latter.

% =========================================================
\section{Conclusions}

In conclusion, the spectral distribution of free electrons after interaction with illuminated atoms reveals energy gain and loss features associated with transitions between Floquet states of the system, emerging at transition energies that undergo dynamical Stark shifts as the light intensity increases. These features add up to those produced by a net number of photon exchanges between the electron and the external light, mediated by the near field associated with optical scattering by the atom, similar to what is observed in PINEM experiments for linearly responding samples \cite{BFZ09,paper151}. The evolution of Floquet transitions and PINEM-like sidebands follow a complex dynamics as a function of photon frequency, including the presence of multiple avoided crossings. This type of free electron-atom scattering could be explored for e-beams traversing diluted atomic gases, while we expect excitons in two-dimensional materials, which are currently a focus of attention in the electron microscopy community \cite{TLM15}, to also display a Floquet dynamics similar to what we describe here under intense laser irradiation. Both of these configurations are feasible using currently available ultrafast electron microscopes, where optical modulation of the electron wave function could dramatically affect the subsequent interaction with Floquet systems driven by phase-locked illumination.

% --- acknowledgments -------------------------------------
%\section*{Funding} % ... text ... %---OSA, place funding here and thanks to colleagues below
\section*{ACKNOWLEDGEMENTS} %---ACS---arxiv
%\section*{Acknowledgments} %---OSA
%\begin{acknowledgments} %---APS
This work has been supported in part by the European Research Council (Advanced Grant 789104-eNANO), the European Commission (Horizon 2020 Grants 101017720 FET-Proactive EBEAM and 964591-SMART-electron), the Spanish MINECO (PID2020-112625GB-I00 and Severo Ochoa CEX2019-000910-S), the Catalan CERCA Program, and Fundaci\'{o}s Cellex and Mir-Puig. V.D.G. acknowledges support from the EU (Marie Sk\l{}odowska-Curie Grant 713729). %---ACS---OSA---arxiv
%\end{acknowledgments} %---APS
%\section*{Disclosure} The authors declare no conflicts of interest. %---OSA

% =========================================================
% --- methods/appendix ------------------------------------
% =========================================================
%\section*{METHODS} %---ACS optional
\begin{widetext} %---arxiv
\section*{APPENDIX}
\appendix %---APS---OSA---SI---arxiv optional
%\renewcommand{\thesection}{A} %---SI---arxiv optional
%\renewcommand{\theequation}{A\arabic{equation}} %---SI---arxiv optional
%\setcounter{equation}{0} %---OSA optional

% --- text for methods/appendix ----------------------------
%\noindent{\bf Title Format.} ... text ... %---ACS format for Methods section headers
%\section{title} ... text ... %---APS---SI---arxiv format for Appendix section headers (not for OSA)
%\subsection{title} ... text ... %---APS---OSA---SI---arxiv format for Appendix subsection headers
%\renewcommand{\theequation}{A\arabic{equation}} %---OSA for A, B, ... subsections of Appendix

% =========================================================
% --- SI, acknowledgments, bibliography, etc. -------------
% =========================================================

% --- SI --------------------------------------------------
%\section*{Supplementary Information} %---ACS optional
%The Supporting Information is available free of charge at https://pubs.acs.org/doi/xxx. %---ACS optional

% =========================================================
% --- Floquet theory --------------------------------------
% =========================================================
\section{Floquet theory for an electronic system under external illumination}
\label{App1}

We review the Floquet theory for a system subject to the periodic drive of an external classical monochromatic light field \cite{S1965,F1987}. Considering a set of $N$ electronic states $|a\rangle$ of energies $\hbar\varepsilon_a$, we write the system Hamiltonian as
\begin{align}
\hat{\mathcal{H}}_s(t)=\sum_a \hbar\varepsilon_a|a\rangle\langle a|+\cos(\wL t)\sum_{aa'}\hbar\Om_{aa'}|a\rangle\langle a'|,
\nonumber
\end{align}
where the symmetric coefficients $\Om_{aa'}=\Om_{a'a}$ are optical Rabi frequencies (proportional to the light intensity) and $\wL$ is the light frequency. The Schr\"odinger equation $\hat{\mathcal{H}}_s(t)|\psi(t)\rangle=\ii\hbar|\dot\psi(t)\rangle$ admits $N$ independent steady-state solutions (the so-called Floquet states) of the form \cite{S1965} $|\psi(t)\rangle=\ee^{-\ii\tw_jt}|F_j(t)\rangle$, where the states
\begin{align}
|F_j(t)\rangle=\sum_{al} \ee^{-\ii l\wL t} f_{jal}|a\rangle
\nonumber
\end{align}
have the same periodicity as the external light (with time-independent real coefficients $f_{jal}$) and $\tw_j$ are Floquet frequencies. We can verify that these are indeed solutions of the Schr\"odinger equation by identifying terms with the same time dependence, which leads to
\begin{align}
\tw_j f_{jal} = (\varepsilon_a-l\wL) f_{jal} + \sum_{a'} \Om_{aa'} (f_{ja'l+1}+f_{ja'l-1})/2.
\nonumber
\end{align}
The secular matrix of this system of equations is real and symmetric, so it possesses $N$ real eigenvalues $\tw_j$ satisfying $0\le|\tw_j|\le\wL$, and their corresponding real eigenvectors $f_{jal}$. In addition, $\tw_j+m\wL$ and $f_{jal+m}$ are also eigenvalues and eigenvectors for any integer $m$. After normalization, these solutions must satisfy the orthogonality and completeness relations
\begin{subequations} \label{orthogonalityandcompleteness}
\begin{align}
\sum_{al} f_{jal}f_{j'al+m}=\delta_{jj'}\delta_{m,0} \quad&\rightarrow \quad\langle F_j(t)|F_{j'}(t)\rangle=\delta_{jj'}, \label{orthogonality}\\
\sum_{jm} f_{jal+m}f_{ja'l'+m}=\delta_{aa'}\delta_{ll'} \quad&\rightarrow \quad\sum_j|F_j(t)\rangle\langle F_j(t)|=\sum_a|a\rangle\langle a|. \label{completeness}
\end{align}
\end{subequations}
From the condition that the sum of eigenvalues is equal to the trace of the secular matrix, we find the additional result $\sum_j\tw_j=\sum_a \varepsilon_a$ modulus $\wL$ \cite{S1965}. In the present work, we calculate $\tw_j$ and $f_{jal}$ for atoms with $N=2$ and $N=3$ levels by retaining only $|l|\le l_{\rm max}$ terms and numerically solving the resulting finite system of equations until convergence with increased $l_{\rm max}$ is achieved (typically for $l_{\rm max}<20$).

We introduce dissipation via inelastic jumps between states $a\rightarrow a'$ at rates $\kappa_{aa'}$ by writing the Lindblad master equation of motion
\begin{align}
\ii\hbar\dot{\hat\rho}_s(t)=\left[\hat{\mathcal{H}}_s(t),\hat\rho_s(t)\right]+\ii\hbar\mathcal{L}[\hat\rho_s(t)]
\nonumber
\end{align}
for the density matrix $\hat\rho_s$ of the illuminated system, where
\begin{align}
\mathcal{L}[\hat\rho_s(t)]=\sum_{aa'}\frac{\kappa_{aa'}}{2}\bigg[2|a'\rangle\langle a|\hat\rho_s(t)|a\rangle\langle a'|-|a\rangle\langle a|\hat\rho_s(t)-\hat\rho_s(t)|a\rangle\langle a|\bigg].
\label{Lindbladian}
\end{align}
Projecting on electronic states as $\hat\rho_s(t)=\sum_{aa'}\rrho_{aa'}(t)\,|a\rangle\langle a'|$, we find the equation
\begin{align}
\ii\dot\rrho_{aa'}(t)=\left(\varepsilon_a-\varepsilon_{a'}\right)\rrho_{aa'}(t)&+\cos(\wL t)\sum_{a''}\left[\Om_{aa''}\rrho_{a''a'}(t)-\rrho_{aa''}(t)\Om_{a''a'}\right] \label{rhosecular}\\
&+\ii\sum_{a''}\left[\delta_{aa'}\kappa_{a''a}\rrho_{a''a''}(t)-\left(\kappa_{aa''}+\kappa_{a'a''}\right)\rrho_{aa'}(t)/2\right]
\nonumber
\end{align}
for the expansion coefficients $\rrho_{aa'}(t)$. Due to the presence of decay, there is just a single steady-state solution for the optically driven system, in which $\rrho_{aa'}(t)$ are periodic functions of time with the same period as the incident light, so they can be expanded as
\begin{align}
\rrho_{aa'}(t)=\sum_l\ee^{-\ii l \wL t}\rrho_{aa'l}
\nonumber
\end{align}
in terms of time-independent coefficients $\rrho_{aa'l}$. We also calculate these coefficients numerically and find excellent agreement when solving Eq.\ (\ref{rhosecular}) either in the time domain (after propagation over a long time $\gg1/{\rm min}\{\kappa_{aa'}\}$ to ensure the exponential attenuation of any signature of the initial conditions) or as a linear system of equations separated in different $\ee^{-\ii l \wL t}$ components (from which the steady-state solution emerges as the only eigenstate with a vanishing eigenvalue). Using the completeness of Floquet states [Eqs.\ (\ref{completeness})], it is convenient to represent the resulting steady-state density matrix as
\begin{align}
\hat\rho_s(t)=\sum_{jj'}\trho_{jj'}(t)\,|F_j(t)\rangle\langle F_{j'}(t)|,
\label{rhos}
\end{align}
where the coefficients $\trho_{jj'}(t)=\sum_l\trho_{jj'l}\ee^{-\ii l \wL t}$, with $\trho_{jj'l_0}=\sum_{aa'll'}\rrho_{aa'l_0+l-l'}f_{jal}f_{j'a'l'}$, again display the same temporal periodicity as the applied light field. Incidentally, the Hermiticity of $\hat\rho_s$ leads to the conditions $\rrho_{aa'l}^*=\rrho_{a'a,-l}$ and $\trho_{jj'l}^*=\trho_{j'j,-l}$.

% =========================================================
% --- EELS ------------------------------------------------
% =========================================================
\section{Electron-beam interaction with an illuminated system}
\label{App2}

We consider a collimated electron interacting with an illuminated sample under the following approximations:
\begin{enumerate}[\it\red {\rm(}i\rm{)} ---]
\item Inelastic transitions in the sample occur at a sufficiently small rate as to allow us to neglect them during the electron-sample interaction time.
\item The electron-sample coupling is sufficiently weak as to be describable at the lowest-order level of perturbation theory.
\item The energy width of the incident electron is small compared to the photon energy.
\item The electron velocity vector remains nearly constant during the interaction time (nonrecoil approximation), so the transferred energy only depends on the change of electron momentum along the direction of motion.
\end{enumerate}
Assumption {\it \red {\rm(}i\rm{)}} depends on the choice of sample, but can be easily satisfied for electron pulses of sub-picosecond duration and sample inelastic processes dominated by coupling to radiation, while approximations {\it \red {\rm(}ii\rm{)}}-{\it \red {\rm(}iv\rm{)}} are generally applicable when focusing on samples that host optical resonances in the visible regime and the electron beam (e-beam) is prepared under typical conditions in electron microscopes \cite{paper371}. The density matrix $\hat\rho(t)$ of the entire system satisfies the equation of motion 
\begin{align}
\ii\hbar\dot{\hat\rho}(t)=\left[\hat{\mathcal{H}}_s(t)+\hat{\mathcal{H}}_e+\hat{\mathcal{H}}_{\rm e-s},\;\hat\rho(t)\right]+\ii\hbar\mathcal{L}[\hat\rho(t)],
\label{dtrho}
\end{align}
where $\hat{\mathcal{H}}_e$ is the free-electron Hamiltonian, $\hat{\mathcal{H}}_{e-s}$ describes the electron-sample interaction, and the Lindbladian $\mathcal{L}[\hat\rho(t)]$ acts on the sample degrees of freedom as prescribed by Eq.\ (\ref{Lindbladian}). We represent the electron in terms of states $|q\rangle$ of well-defined longitudinal momentum $\hbar q$ and energy $\hbar\varepsilon_q$, so the free-electron Hamiltonian reduces to $\hat{\mathcal{H}}_e=\sum_q \hbar\varepsilon_q |q\rangle\langle q|$, while the interaction Hamiltonian can be written as
\begin{align}
\hat{\mathcal{H}}_{\rm e-s}=\sum_{aa'qq'}\db_{aa'}\cdot\gb_{q'-q}|aq\rangle\langle a'q'|,
\label{Hes}
\end{align}
where the coupling coefficients are explicitly shown to depend on the momentum difference via the vectors $\gb_{q'-q}$ times the transition dipoles $\db_{aa'}$ (see Ref.\ \cite{paper221}).

We now proceed by expanding the density matrix in a complete basis set of Floquet and electron states as
\begin{align}
\hat\rho(t)=\sum_{jqj'q'}\alpha_{jqj'q'}(t)\,\ee^{\ii(\tw_{j'}-\tw_j+\varepsilon_{q'}-\varepsilon_q)t}\,|F_j(t),q\rangle\langle F_{j'}(t),q'|,
\label{rho}
\end{align}
where we incorporate the Floquet frequencies $\tw_j$ in such a way that the coefficients $\alpha_{jqj'q'}(t)$ are time-independent outside the interaction window. In particular, prior to interaction, the density matrix reduces to
\begin{align}
\hat\rho^{(0)}(t)=\sum_{jqj'q'}\alpha_q^0\alpha_{q'}^{0*}\,\trho_{jj'}(t)\,\ee^{\ii(\varepsilon_{q'}-\varepsilon_q)t}\,|F_j(t),q\rangle\langle F_{j'}(t),q'|,
\nonumber
\end{align}
which is constructed from Eq.\ (\ref{rhos}) by introducing the time-independent coefficients $\alpha_q^0$ of the incident electron wave function $\sum_q \alpha_q^0\,\ee^{-\ii\varepsilon_qt}|q\rangle$. Therefore, at zeroth order of interaction, we have the density matrix coefficients
\begin{align}
\alpha^{(0)}_{jqj'q'}(t)=\alpha_q^0\alpha_{q'}^{0*}\,\trho_{jj'}(t)\,\ee^{\ii(\tw_j-\tw_{j'})t},
\label{alpha0}
\end{align}
where $\trho_{jj'}(t)$ describes the steady-state solution of the illuminated system [Eq.\ (\ref{rhos})].

According to item {\it \red {\rm(}i\rm{)}} in the above list of assumptions, we can neglect the Lindbladian in Eq.\ (\ref{dtrho}) to propagate the density matrix along the interaction region. Then, inserting Eq.\ (\ref{rho}) into Eq.\ (\ref{dtrho}) and using the orthogonality relation of Eq.\ (\ref{orthogonality}), we can construct the interaction series $\alpha_{jqj'q'}=\sum_n \alpha^{(n)}_{jqj'q'}$, whose terms satisfy the recursion relation
\begin{align}
\ii\hbar\dot\alpha^{(n)}_{jqj'q'}(t)=\sum_{j''q''}\bigg[
&\alpha^{(n-1)}_{j''q''j'q'}(t) \,\langle F_j(t)q|\hat{\mathcal{H}}_{e-s}|F_{j''}(t)q''\rangle \,\ee^{\ii(\tw_j-\tw_{j''}+\varepsilon_q-\varepsilon_{q''})t} \label{recn}\\
-&\alpha^{(n-1)}_{jqj''q''}(t) \,\langle F_{j''}(t)q''|\hat{\mathcal{H}}_{e-s}|F_{j'}(t)q'\rangle \,\ee^{\ii(\tw_{j''}-\tw_{j'}+\varepsilon_{q''}-\varepsilon_{q'})t}
\bigg]. \nonumber
\end{align}
The lowest-order of perturbation theory [approximation {\it \red {\rm(}ii\rm{)}}] that produces changes in the electron energy distribution (i.e., modifications in the diagonal elements of the density matrix) is $n=2$, for which the post-interaction inelastic electron distribution (i.e., the trace over sample degrees of freedom at infinite time) reads
\begin{align}
P_q&=\sum_j \alpha^{(2)}_{jqjq}(\infty)
=\frac{2}{\hbar}\sum_{jj'q'}\int_{-\infty}^\infty dt\;{\rm Im}\left\{\alpha^{(1)}_{j'q'jq}(t)\,\langle F_j(t)q|\hat{\mathcal{H}}_{e-s}|F_{j'}(t)q'\rangle \,\ee^{\ii(\tw_j-\tw_{j'}+\varepsilon_q-\varepsilon_{q'})t}\right\}, \label{Pq}
\end{align}
where the rightmost expression is obtained by integrating Eq.\ (\ref{recn}) and applying the Hermiticity property $\big[\alpha^{(1)}_{j'q'jq}(t)\big]^*=\alpha^{(1)}_{jqj'q'}(t)$. At this point, it is convenient to rewrite the matrix elements in Eq.\ (\ref{Pq}) as
\begin{align}
\langle F_j(t)q|\hat{\mathcal{H}}_{e-s}|F_{j'}(t)q'\rangle\equiv\hbar\sum_l N_{jj'l,q'-q}\,\ee^{-\ii l\wL t},
\label{FHF}
\end{align}
where we introduce the coefficients
\begin{align}
N_{jj'l,p}=\frac{1}{\hbar}\sum_{aa'l'}f_{jal'}\,f_{j'a'l+l'}\;\db_{aa'}\cdot\gb_p,
\label{Njjlp}
\end{align}
which have units of frequency and satisfy the symmetry relation
\begin{align}
N^*_{jj'l,p}=N_{j'j,-l,-p}.
\label{symm}
\end{align}
Also, we evaluate the coefficient $\alpha^{(1)}_{j'q'jq}(t)$ needed in Eq.\ (\ref{Pq}) by inserting Eq.\ (\ref{alpha0}) into Eq.\ (\ref{recn}) and setting $n=1$. In the process, we encounter the sums $\sum_{j''}\trho_{j''j'}(t)\,\langle F_j(t)q|\hat{\mathcal{H}}_{e-s}|F_{j''}(t)q'\rangle\equiv\hbar\sum_l M_{jj'l,q'-q}\ee^{-\ii l\wL t}$, where we implicitly define the time-independent coefficients
\begin{align}
M_{jj'l,p}
&=\frac{1}{\hbar}\sum_{j''}\sum_{a a' l'' l'''}
\trho_{j''j'l-l'}\,f_{jal''}\,f_{j''a'l'+l''}\;\db_{aa'}\cdot\gb_p \nonumber\\
&=\sum_{j''l'}
\trho_{j''j'l-l'}\,N_{jj''l',p}.
\label{Mjjlp}
\end{align}
This allows us to write
\begin{align}
\dot\alpha^{(1)}_{jqj'q'}(t)=-\ii\sum_{lq''}\bigg[
&\alpha_{q''}^0\alpha_{q'}^{0*}\, M_{jj'l,q''-q}
\,\ee^{\ii(\tw_j-\tw_{j'}+\varepsilon_q-\varepsilon_{q''}-l\wL)t}
-\alpha_q^0\alpha_{q''}^{0*}\, M^*_{j'jl,q''-q'}
\,\ee^{\ii(\tw_j-\tw_{j'}+\varepsilon_{q''}-\varepsilon_{q'}+l\wL)t}
\bigg], \nonumber
\end{align}
which can be readily integrated to yield
\begin{align}
\alpha^{(1)}_{jqj'q'}(t)=-\sum_{lq''}\bigg[
&\alpha_{q''}^0\alpha_{q'}^{0*}\, M_{jj'l,q''-q}
\,\frac{\ee^{\ii(\tw_j-\tw_{j'}+\varepsilon_q-\varepsilon_{q''}-l\wL)t}}{\tw_j-\tw_{j'}+\varepsilon_q-\varepsilon_{q''}-l\wL-\ii0^+} \nonumber\\
-&\alpha_q^0\alpha_{q''}^{0*}\, M^*_{j'jl,q''-q'}
\,\frac{\ee^{\ii(\tw_j-\tw_{j'}+\varepsilon_{q''}-\varepsilon_{q'}+l\wL)t}}{\tw_j-\tw_{j'}+\varepsilon_{q''}-\varepsilon_{q'}+l\wL-\ii0^+}
\bigg]. \nonumber
\end{align}
Now, inserting this result into Eq.\ (\ref{Pq}) and using Eqs.\ (\ref{FHF}) and (\ref{symm}), we find
\begin{align}
P_q=-4\pi\sum_{jj'll'}\sum_{q'q''}{\rm Im}\bigg\{
&\alpha_{q''}^0\alpha_{q}^{0*}\, M_{j'jl,q''-q'}N_{jj'l',q'-q}
\,\frac{\delta\big[\varepsilon_q-\varepsilon_{q''}-(l+l')\wL\big]}{\tw_{j'}-\tw_j+\varepsilon_{q'}-\varepsilon_{q''}-l\wL-\ii0^+} \nonumber\\
-&\alpha_{q'}^0\alpha_{q''}^{0*}\, M^*_{jj'l,q''-q}N_{jj'l',q'-q}
\,\frac{\delta\big[\varepsilon_{q''}-\varepsilon_{q'}+(l-l')\wL\big]}{\tw_{j'}-\tw_j+\varepsilon_{q''}-\varepsilon_q+l\wL-\ii0^+} \bigg\}. \nonumber
\end{align}
Assumption {\it \red {\rm(}iii\rm{)}} from the above list implies that $\delta(\varepsilon_q-\varepsilon_{q'}-l\wL)\alpha_q^0\alpha_{q'}^{0*}$ is negligible unless $l=0$ (i.e., displacements of the incident electron wavepacket by multiples of the photon energy produce nonoverlapping peaks). Then, the $l'$ sum is only contributed by $l'=-l$ and $l'=l$ terms in the first and second lines of the above expression, respectively, so ta it simplifies to
\begin{align}
P_q=-\frac{2L}{v}\sum_{jj'l}\sum_{q'}\,{\rm Im}\bigg\{
&\left|\alpha_{q}^0\right|^2\, M^*_{jj'l,q-q'} N_{jj',l,q-q'}
\,\frac{1}{\tw_{j'}-\tw_{j}+(q-q')v+l\wL-\ii0^+} \label{Pq1}\\
-&\left|\alpha_{q'}^0\right|^2\, M^*_{jj'l,q'-q} N_{jj'l,q'-q}
\,\frac{1}{\tw_{j'}-\tw_j+(q'-q)v+l\wL-\ii0^+} \bigg\}, \nonumber
\end{align}
where $v$ is the electron velocity. To obtain this result, we have further invoked the nonrecoil approximation \cite{paper371} [item {\it \red {\rm(}iv\rm{)}} from the above list, which allows us to write $\varepsilon_q-\varepsilon_{q'}\approx (q-q')v$], converted the sum over $q$ into an integral by means of the prescription $\sum_q\rightarrow (L/2\pi)\int dq$, where $L$ is the quantization length along the e-beam direction, and used the symmetry relation in Eq.\ (\ref{symm}).

Reassuringly, the $q\leftrightarrow q'$ symmetry between the two lines in Eq.\ (\ref{Pq1}) readily leads to conservation of the total electron probability (i.e., $\sum_q P_q=0$). In particular, the first line in Eq.\ (\ref{Pq1}) describes scattering terms in which the final wave vector $q$ is contained within the zero-loss peak (ZLP) of the incident electron, as imposed by the $\left|\alpha_q^0\right|^2$ factor inside the integrand, and consequently, this contribution must be negative, implying a reduction of the ZLP at the expense of the inelastic scattering probability produced by the second line. In addition, the resulting energy distribution must consist of peaks with a similar energy width as the ZLP because this is large compared with the width of the sample resonances [assumption {\it \red {\rm(}i\rm{)}}] and we are considering monochromatic light.

Obviously, the initial electron probability must be normalized, so we have $\sum_q\left|\alpha_q^0\right|^2=1$. Considering a central wave vector $q_0$ in the incident electron wavepacket and using the nonrecoil approximation [assumption {\it \red {\rm(}iv\rm{)}}] to associate each $q$ component with an {\it energy loss} \cite{Floquetqtow} $\hbar\omega=-\hbar(q-q_0)v$ relative to the central energy $\hbar\varepsilon_{q_0}$, we can recast the normalization condition as $\int_{-\infty}^\infty d\omega \,\Gamma_{\rm ZLP}(\omega)=1$ in terms of the incident electron energy distribution $\Gamma_{\rm ZLP}(\omega)=(L/2\pi v)\big|\alpha_{q_0-\omega/v}^0\big|^2$, where we have used again the prescription that transforms the $q$ sum into an integral.

From the above considerations, the spectral probability of transmitted electrons can be approximated as
\begin{align}
\Gamma_{\rm EELS}(\omega)\approx P_{\rm ZLP}\Gamma_{\rm ZLP}(\omega)+{\sum_{jj'l}}' P_{jj'l}\,\Gamma_{\rm ZLP}(\omega-\omega_{jj'l}), \label{gw}
\end{align}
which consists of a series of peaks at energies $\hbar\omega_{jj'l}\equiv\hbar(\tw_j-\tw_{j'}+l\wL)$ with probabilities
\begin{align}
P_{jj'l}&=\frac{L^2}{\pi v}\int dp\,{\rm Im}\bigg\{
M^*_{jj',-l,p} N_{jj',-l,p}
\,\frac{1}{pv-\omega_{jj'l}-\ii0^+} \bigg\} \nonumber\\
&\approx\frac{L^2}{v^2}\;{\rm Re}\left\{
M^*_{jj',-l,\omega_{jj'l}/v} N_{jj',-l,\omega_{jj'l}/v}\right\}. \label{Pjjl}
\end{align}
The sum in Eq.\ (\ref{gw}) is restricted to terms in which $j\neq j'$ or $l\neq0$ (i.e., separated from the ZLP), whereas, in virtue of the conservation of the total probability (see above), the ZLP term must have a probability $P_{\rm ZLP}=1-\sum_{jj'l}' P_{jj'l}$. The second line of Eq.\ (\ref{Pjjl}) is obtained by neglecting a contribution proportional to ${\rm Im}\left\{
M^*_{jj',-l,\omega_{jj'l}/v} N_{jj',-l,\omega_{jj'l}/v}\right\}$, which we find to be indeed an excellent approximation in our calculations under the assumption {\it \red {\rm(}i\rm{)}} of the above list. We argue here that the coefficients $\trho_{jj'l}$, entering this term through Eq.\ (\ref{Mjjlp}), have an imaginary part that vanishes in the limit of small decay rates; also, the approximation of neglecting the Lindbladian during the interaction interval can introduce small corrections that are proportional to the decay rates, so the bottom-most result in Eq.\ (\ref{Pjjl}) appears to be consistent with assumption {\it \red {\rm(}i\rm{)}}.

For atomic and molecular systems hosting confined electronic modes, the transition dipoles $\db_{aa'}$ [see Eq.\ (\ref{Hes})] are real. For simplicity, we take them to be oriented along the impact parameter vector $\hat\Rb_e$ (perpendicular to the e-beam direction), so the interaction only depends on the component $\hat\Rb_e\cdot\gb_{\omega_{jj'l}/v}=-(2e/L)(|\omega_{jj'l}|/v\gamma)K_1(|\omega_{jj'l}|R_e/v\gamma)$ of the electron coupling vectors \cite{paper221}, where $\gamma=1/\sqrt{1-v^2/c^2}$ is the Lorentz factor. This expression is real and independent of the transition energy for small beam-sample separation $R_e\ll v\gamma/|\omega_{jj'l}|$ (i.e., $\hat\Rb_e\cdot\gb_{\omega_{jj'l}/v}\approx-2e/R_eL$). Incidentally, the principal value contribution of $1/(pv-\omega_{jj'l})$ to the integral in Eq.\ (\ref{Pjjl}) vanishes in this limit because $M^*_{jj',-l,p} N_{jj',-l,p}$ is independent of $p$, and therefore, the approximated expression in the second line of that equation becomes exact. This allows us to present results in the main text normalized to the EELS probability in the absence of illumination. In particular, for a non-illuminated system (Rabi frequencies $\Omega_{aa'}=0$) initially prepared in the ground state ($a=0$), we have an inelastic feature corresponding to an electron energy loss $\hbar\omega_0=\hbar(\varepsilon_1-\varepsilon_0)$ (transition to $a=1$) with probability $P^0=(L^2/\hbar^2v^2)\big|\db_{10}\cdot\gb_{\omega_0/v}\big|^2$. Under these conditions, from Eqs.\ (\ref{Njjlp}), (\ref{Mjjlp}), and (\ref{Pjjl}), the normalized probability of peak $jj'l$ emerging at an electron energy loss $\hbar\omega_{jj'l}\equiv\hbar(\tw_j-\tw_{j'}+l\wL)$ reduces to
\begin{align}
\frac{P_{jj'l}}{P^0}=I_{jj'l}\sum_{j''l'} I_{jj''l'}\;
{\rm Re}\{\trho_{j''j',l'-l}\}, \nonumber
\end{align}
where
\begin{align}
I_{jj'l}=\sum_{aa'l'}f_{jal'}\,f_{j'a',l'-l}\frac{d_{aa'}}{d_{10}}. \nonumber
\end{align}
These are the expressions actually used to obtain the results of the main text, where the coefficients $f_{jal}$ and $\trho_{jj'l}$, as well as the frequencies $\tw_j$, are calculated as explained in Sec.\ \ref{App1}. For a two-level system, the ratio between transition dipoles in $I_{jj'l}$ is 1. For three-level atoms, we consider a mixture of atomic transitions that couple to the incident light, to the electron, or to both of them, depending on the dipole orientations. A practical realization of this idea could be implemented by playing with the light polarization and exploiting the fact that coupling to the electron is forbidden if the transition dipole is perpendicular to both $\Rb_e$ and $\vb$.

\clearpage %--- optional
\pagebreak \onecolumngrid \section*{SUPPLEMENTARY FIGURE} %---SI---arxiv optional

%Figure S1 ------------------------------------------------
\begin{figure*}[h] \label{FigS1}
\begin{centering} \includegraphics[width=0.9\textwidth]{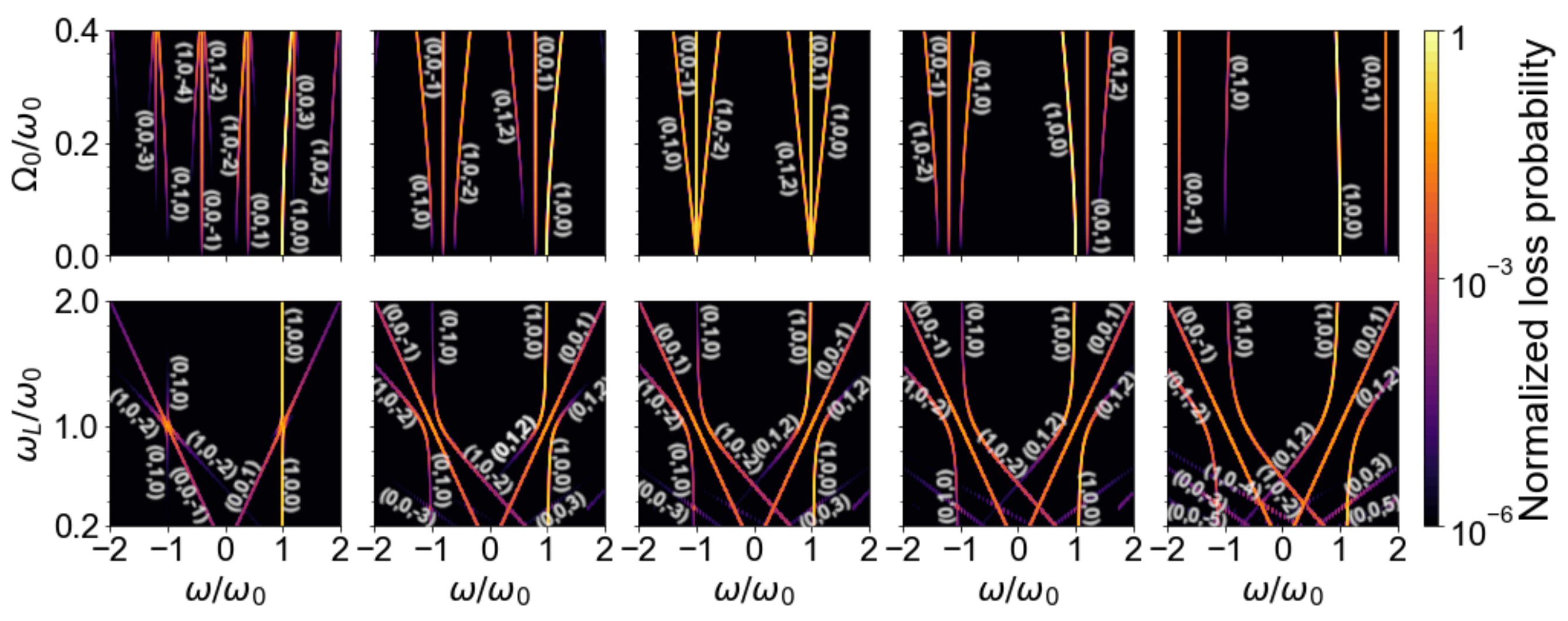} \par\end{centering}
\caption{Same as Fig.\ \ref{Fig2}, but with the spectral features labeled by their corresponding $(jj'l)$ indices.}
\end{figure*}

\end{widetext} %---arxiv

% --- bibliography (adapt file path as appropriate) -------
%\bibliographystyle{apsrev} %---APS---SI---arxiv %---comment out for longbibliography
%\bibliography{../../../bibtex/refsL.bib} %---APS---OSA---SI---arxiv with lower-case title format
%\bibliography{../../../bibtex/refsU.bib} %---ACS with upper-case title format

\begin{thebibliography}{77}
\expandafter\ifx\csname natexlab\endcsname\relax\def\natexlab#1{#1}\fi
\expandafter\ifx\csname bibnamefont\endcsname\relax
  \def\bibnamefont#1{#1}\fi
\expandafter\ifx\csname bibfnamefont\endcsname\relax
  \def\bibfnamefont#1{#1}\fi
\expandafter\ifx\csname citenamefont\endcsname\relax
  \def\citenamefont#1{#1}\fi
\expandafter\ifx\csname url\endcsname\relax
  \def\url#1{\texttt{#1}}\fi
\expandafter\ifx\csname urlprefix\endcsname\relax\def\urlprefix{URL }\fi
\providecommand{\bibinfo}[2]{#2}
\providecommand{\eprint}[2][]{\url{#2}}

\bibitem[{\citenamefont{Batson et~al.}(2002)\citenamefont{Batson, Dellby, and
  Krivanek}}]{BDK02}
\bibinfo{author}{\bibfnamefont{P.~E.} \bibnamefont{Batson}},
  \bibinfo{author}{\bibfnamefont{N.}~\bibnamefont{Dellby}}, \bibnamefont{and}
  \bibinfo{author}{\bibfnamefont{O.~L.} \bibnamefont{Krivanek}},
  \bibinfo{journal}{Nature} \textbf{\bibinfo{volume}{418}},
  \bibinfo{pages}{617} (\bibinfo{year}{2002}).

\bibitem[{\citenamefont{Krivanek et~al.}(2014)\citenamefont{Krivanek, Lovejoy,
  Dellby, Aoki, Carpenter, Rez, Soignard, Zhu, Batson, Lagos et~al.}}]{KLD14}
\bibinfo{author}{\bibfnamefont{O.~L.} \bibnamefont{Krivanek}},
  \bibinfo{author}{\bibfnamefont{T.~C.} \bibnamefont{Lovejoy}},
  \bibinfo{author}{\bibfnamefont{N.}~\bibnamefont{Dellby}},
  \bibinfo{author}{\bibfnamefont{T.}~\bibnamefont{Aoki}},
  \bibinfo{author}{\bibfnamefont{R.~W.} \bibnamefont{Carpenter}},
  \bibinfo{author}{\bibfnamefont{P.}~\bibnamefont{Rez}},
  \bibinfo{author}{\bibfnamefont{E.}~\bibnamefont{Soignard}},
  \bibinfo{author}{\bibfnamefont{J.}~\bibnamefont{Zhu}},
  \bibinfo{author}{\bibfnamefont{P.~E.} \bibnamefont{Batson}},
  \bibinfo{author}{\bibfnamefont{M.~J.} \bibnamefont{Lagos}},
  \bibnamefont{et~al.}, \bibinfo{journal}{Nature}
  \textbf{\bibinfo{volume}{514}}, \bibinfo{pages}{209} (\bibinfo{year}{2014}).

\bibitem[{\citenamefont{Krivanek et~al.}(2019)\citenamefont{Krivanek, Dellby,
  Hachtel, Idrobo, Hotz, Plotkin-Swing, Bacon, Bleloch, Corbin, Hoffman
  et~al.}}]{KDH19}
\bibinfo{author}{\bibfnamefont{O.~L.} \bibnamefont{Krivanek}},
  \bibinfo{author}{\bibfnamefont{N.}~\bibnamefont{Dellby}},
  \bibinfo{author}{\bibfnamefont{J.~A.} \bibnamefont{Hachtel}},
  \bibinfo{author}{\bibfnamefont{J.-C.} \bibnamefont{Idrobo}},
  \bibinfo{author}{\bibfnamefont{M.~T.} \bibnamefont{Hotz}},
  \bibinfo{author}{\bibfnamefont{B.}~\bibnamefont{Plotkin-Swing}},
  \bibinfo{author}{\bibfnamefont{N.~J.} \bibnamefont{Bacon}},
  \bibinfo{author}{\bibfnamefont{A.~L.} \bibnamefont{Bleloch}},
  \bibinfo{author}{\bibfnamefont{G.~J.} \bibnamefont{Corbin}},
  \bibinfo{author}{\bibfnamefont{M.~V.} \bibnamefont{Hoffman}},
  \bibnamefont{et~al.}, \bibinfo{journal}{Ultramicroscopy}
  \textbf{\bibinfo{volume}{203}}, \bibinfo{pages}{60} (\bibinfo{year}{2019}).

\bibitem[{\citenamefont{Lagos et~al.}(2017)\citenamefont{Lagos, Tr\"ugler,
  Hohenester, and Batson}}]{LTH17}
\bibinfo{author}{\bibfnamefont{M.~J.} \bibnamefont{Lagos}},
  \bibinfo{author}{\bibfnamefont{A.}~\bibnamefont{Tr\"ugler}},
  \bibinfo{author}{\bibfnamefont{U.}~\bibnamefont{Hohenester}},
  \bibnamefont{and} \bibinfo{author}{\bibfnamefont{P.~E.}
  \bibnamefont{Batson}}, \bibinfo{journal}{Nature}
  \textbf{\bibinfo{volume}{543}}, \bibinfo{pages}{529} (\bibinfo{year}{2017}).

\bibitem[{\citenamefont{Lagos and Batson}(2018)}]{LB18}
\bibinfo{author}{\bibfnamefont{M.~J.} \bibnamefont{Lagos}} \bibnamefont{and}
  \bibinfo{author}{\bibfnamefont{P.~E.} \bibnamefont{Batson}},
  \bibinfo{journal}{Nano\ Lett.} \textbf{\bibinfo{volume}{18}},
  \bibinfo{pages}{4556} (\bibinfo{year}{2018}).

\bibitem[{\citenamefont{Hage et~al.}(2018)\citenamefont{Hage, Nicholls, Yates,
  McCulloch, Lovejoy, Dellby, Krivanek, Refson, and Ramasse}}]{HNY18}
\bibinfo{author}{\bibfnamefont{F.~S.} \bibnamefont{Hage}},
  \bibinfo{author}{\bibfnamefont{R.~J.} \bibnamefont{Nicholls}},
  \bibinfo{author}{\bibfnamefont{J.~R.} \bibnamefont{Yates}},
  \bibinfo{author}{\bibfnamefont{D.~G.} \bibnamefont{McCulloch}},
  \bibinfo{author}{\bibfnamefont{T.~C.} \bibnamefont{Lovejoy}},
  \bibinfo{author}{\bibfnamefont{N.}~\bibnamefont{Dellby}},
  \bibinfo{author}{\bibfnamefont{O.~L.} \bibnamefont{Krivanek}},
  \bibinfo{author}{\bibfnamefont{K.}~\bibnamefont{Refson}}, \bibnamefont{and}
  \bibinfo{author}{\bibfnamefont{Q.~M.} \bibnamefont{Ramasse}},
  \bibinfo{journal}{Sci.\ Adv.} \textbf{\bibinfo{volume}{4}},
  \bibinfo{pages}{eaar7495} (\bibinfo{year}{2018}).

\bibitem[{\citenamefont{Hachtel et~al.}(2019)\citenamefont{Hachtel, Huang,
  Popovs, Jansone-Popova, Keum, Jakowski, Lovejoy, Dellby, Krivanek, and
  Idrobo}}]{HHP19}
\bibinfo{author}{\bibfnamefont{J.~A.} \bibnamefont{Hachtel}},
  \bibinfo{author}{\bibfnamefont{J.}~\bibnamefont{Huang}},
  \bibinfo{author}{\bibfnamefont{I.}~\bibnamefont{Popovs}},
  \bibinfo{author}{\bibfnamefont{S.}~\bibnamefont{Jansone-Popova}},
  \bibinfo{author}{\bibfnamefont{J.~K.} \bibnamefont{Keum}},
  \bibinfo{author}{\bibfnamefont{J.}~\bibnamefont{Jakowski}},
  \bibinfo{author}{\bibfnamefont{T.~C.} \bibnamefont{Lovejoy}},
  \bibinfo{author}{\bibfnamefont{N.}~\bibnamefont{Dellby}},
  \bibinfo{author}{\bibfnamefont{O.~L.} \bibnamefont{Krivanek}},
  \bibnamefont{and} \bibinfo{author}{\bibfnamefont{J.~C.}
  \bibnamefont{Idrobo}}, \bibinfo{journal}{Science}
  \textbf{\bibinfo{volume}{363}}, \bibinfo{pages}{525} (\bibinfo{year}{2019}).

\bibitem[{\citenamefont{Hage et~al.}(2019)\citenamefont{Hage, Kepaptsoglou,
  Ramasse, and Allen}}]{HKR19}
\bibinfo{author}{\bibfnamefont{F.~S.} \bibnamefont{Hage}},
  \bibinfo{author}{\bibfnamefont{D.~M.} \bibnamefont{Kepaptsoglou}},
  \bibinfo{author}{\bibfnamefont{Q.~M.} \bibnamefont{Ramasse}},
  \bibnamefont{and} \bibinfo{author}{\bibfnamefont{L.~J.} \bibnamefont{Allen}},
  \bibinfo{journal}{Phys.\ Rev.\ Lett.} \textbf{\bibinfo{volume}{122}},
  \bibinfo{pages}{016103} (\bibinfo{year}{2019}).

\bibitem[{\citenamefont{Tizei et~al.}(2020)\citenamefont{Tizei, Mkhitaryan,
  {Louren{\c{c}}o-Martins}, Scarabelli, Watanabe, Taniguchi, Tenc{\'{e}},
  Blazit, Li, Gloter et~al.}}]{paper342}
\bibinfo{author}{\bibfnamefont{L.~H.~G.} \bibnamefont{Tizei}},
  \bibinfo{author}{\bibfnamefont{V.}~\bibnamefont{Mkhitaryan}},
  \bibinfo{author}{\bibfnamefont{H.}~\bibnamefont{{Louren{\c{c}}o-Martins}}},
  \bibinfo{author}{\bibfnamefont{L.}~\bibnamefont{Scarabelli}},
  \bibinfo{author}{\bibfnamefont{K.}~\bibnamefont{Watanabe}},
  \bibinfo{author}{\bibfnamefont{T.}~\bibnamefont{Taniguchi}},
  \bibinfo{author}{\bibfnamefont{M.}~\bibnamefont{Tenc{\'{e}}}},
  \bibinfo{author}{\bibfnamefont{J.~D.} \bibnamefont{Blazit}},
  \bibinfo{author}{\bibfnamefont{X.}~\bibnamefont{Li}},
  \bibinfo{author}{\bibfnamefont{A.}~\bibnamefont{Gloter}},
  \bibnamefont{et~al.}, \bibinfo{journal}{Nano\ Lett.}
  \textbf{\bibinfo{volume}{20}}, \bibinfo{pages}{2973} (\bibinfo{year}{2020}).

\bibitem[{\citenamefont{Senga et~al.}(2019)\citenamefont{Senga, Suenaga,
  Barone, Morishita, Mauri, and Pichler}}]{SSB19}
\bibinfo{author}{\bibfnamefont{R.}~\bibnamefont{Senga}},
  \bibinfo{author}{\bibfnamefont{K.}~\bibnamefont{Suenaga}},
  \bibinfo{author}{\bibfnamefont{P.}~\bibnamefont{Barone}},
  \bibinfo{author}{\bibfnamefont{S.}~\bibnamefont{Morishita}},
  \bibinfo{author}{\bibfnamefont{F.}~\bibnamefont{Mauri}}, \bibnamefont{and}
  \bibinfo{author}{\bibfnamefont{T.}~\bibnamefont{Pichler}},
  \bibinfo{journal}{Nature} pp. \bibinfo{pages}{247--250}
  (\bibinfo{year}{2019}).

\bibitem[{\citenamefont{Hage et~al.}(2020)\citenamefont{Hage, Radtke,
  Kepaptsoglou, Lazzeri, and Ramasse}}]{HRK20}
\bibinfo{author}{\bibfnamefont{F.~S.} \bibnamefont{Hage}},
  \bibinfo{author}{\bibfnamefont{G.}~\bibnamefont{Radtke}},
  \bibinfo{author}{\bibfnamefont{D.~M.} \bibnamefont{Kepaptsoglou}},
  \bibinfo{author}{\bibfnamefont{M.}~\bibnamefont{Lazzeri}}, \bibnamefont{and}
  \bibinfo{author}{\bibfnamefont{Q.~M.} \bibnamefont{Ramasse}},
  \bibinfo{journal}{Science} \textbf{\bibinfo{volume}{367}},
  \bibinfo{pages}{1124} (\bibinfo{year}{2020}).

\bibitem[{\citenamefont{Yan et~al.}(2021)\citenamefont{Yan, Liu, Gadre, Gu,
  Aoki, Lovejoy, Dellby, Krivanek, Schlom, Wu et~al.}}]{YLG21}
\bibinfo{author}{\bibfnamefont{X.}~\bibnamefont{Yan}},
  \bibinfo{author}{\bibfnamefont{C.}~\bibnamefont{Liu}},
  \bibinfo{author}{\bibfnamefont{C.~A.} \bibnamefont{Gadre}},
  \bibinfo{author}{\bibfnamefont{L.}~\bibnamefont{Gu}},
  \bibinfo{author}{\bibfnamefont{T.}~\bibnamefont{Aoki}},
  \bibinfo{author}{\bibfnamefont{T.~C.} \bibnamefont{Lovejoy}},
  \bibinfo{author}{\bibfnamefont{N.}~\bibnamefont{Dellby}},
  \bibinfo{author}{\bibfnamefont{O.~L.} \bibnamefont{Krivanek}},
  \bibinfo{author}{\bibfnamefont{D.~G.} \bibnamefont{Schlom}},
  \bibinfo{author}{\bibfnamefont{R.}~\bibnamefont{Wu}}, \bibnamefont{et~al.},
  \bibinfo{journal}{Nature} \textbf{\bibinfo{volume}{589}}, \bibinfo{pages}{65}
  (\bibinfo{year}{2021}).

\bibitem[{\citenamefont{Mkhitaryan et~al.}(2021)\citenamefont{Mkhitaryan,
  March, Tseng, Li, Scarabelli, Liz-Marz\'an, Chen, Tizei, St\'ephan, Song
  et~al.}}]{paper369}
\bibinfo{author}{\bibfnamefont{V.}~\bibnamefont{Mkhitaryan}},
  \bibinfo{author}{\bibfnamefont{K.}~\bibnamefont{March}},
  \bibinfo{author}{\bibfnamefont{E.}~\bibnamefont{Tseng}},
  \bibinfo{author}{\bibfnamefont{X.}~\bibnamefont{Li}},
  \bibinfo{author}{\bibfnamefont{L.}~\bibnamefont{Scarabelli}},
  \bibinfo{author}{\bibfnamefont{L.~M.} \bibnamefont{Liz-Marz\'an}},
  \bibinfo{author}{\bibfnamefont{S.-Y.} \bibnamefont{Chen}},
  \bibinfo{author}{\bibfnamefont{L.~H.~G.} \bibnamefont{Tizei}},
  \bibinfo{author}{\bibfnamefont{O.}~\bibnamefont{St\'ephan}},
  \bibinfo{author}{\bibfnamefont{J.-M.} \bibnamefont{Song}},
  \bibnamefont{et~al.}, \bibinfo{journal}{Nano\ Lett.}
  \textbf{\bibinfo{volume}{21}}, \bibinfo{pages}{2444} (\bibinfo{year}{2021}).

\bibitem[{\citenamefont{Nellist and Pennycook}(1998)}]{NP98}
\bibinfo{author}{\bibfnamefont{P.~D.} \bibnamefont{Nellist}} \bibnamefont{and}
  \bibinfo{author}{\bibfnamefont{S.~J.} \bibnamefont{Pennycook}},
  \bibinfo{journal}{Phys.\ Rev.\ Lett.} \textbf{\bibinfo{volume}{81}},
  \bibinfo{pages}{4156} (\bibinfo{year}{1998}).

\bibitem[{\citenamefont{Grinolds et~al.}(2006)\citenamefont{Grinolds, Lobastov,
  Weissenrieder, and Zewail}}]{GLW06}
\bibinfo{author}{\bibfnamefont{M.~S.} \bibnamefont{Grinolds}},
  \bibinfo{author}{\bibfnamefont{V.~A.} \bibnamefont{Lobastov}},
  \bibinfo{author}{\bibfnamefont{J.}~\bibnamefont{Weissenrieder}},
  \bibnamefont{and} \bibinfo{author}{\bibfnamefont{A.~H.}
  \bibnamefont{Zewail}}, \bibinfo{journal}{Proc.\ Natl.\ Academ.\ Sci.}
  \textbf{\bibinfo{volume}{103}}, \bibinfo{pages}{18427}
  (\bibinfo{year}{2006}).

\bibitem[{\citenamefont{Barwick et~al.}(2008)\citenamefont{Barwick, Park, Kwon,
  Baskin, and Zewail}}]{BPK08}
\bibinfo{author}{\bibfnamefont{B.}~\bibnamefont{Barwick}},
  \bibinfo{author}{\bibfnamefont{H.~S.} \bibnamefont{Park}},
  \bibinfo{author}{\bibfnamefont{O.~H.} \bibnamefont{Kwon}},
  \bibinfo{author}{\bibfnamefont{J.~S.} \bibnamefont{Baskin}},
  \bibnamefont{and} \bibinfo{author}{\bibfnamefont{A.~H.}
  \bibnamefont{Zewail}}, \bibinfo{journal}{Science}
  \textbf{\bibinfo{volume}{322}}, \bibinfo{pages}{1227} (\bibinfo{year}{2008}).

\bibitem[{\citenamefont{Barwick et~al.}(2009)\citenamefont{Barwick, Flannigan,
  and Zewail}}]{BFZ09}
\bibinfo{author}{\bibfnamefont{B.}~\bibnamefont{Barwick}},
  \bibinfo{author}{\bibfnamefont{D.~J.} \bibnamefont{Flannigan}},
  \bibnamefont{and} \bibinfo{author}{\bibfnamefont{A.~H.}
  \bibnamefont{Zewail}}, \bibinfo{journal}{Nature}
  \textbf{\bibinfo{volume}{462}}, \bibinfo{pages}{902} (\bibinfo{year}{2009}).

\bibitem[{\citenamefont{Aseyev et~al.}(2020)\citenamefont{Aseyev, Ryabov,
  Mironov, and Ischenko}}]{ARM20}
\bibinfo{author}{\bibfnamefont{S.~A.} \bibnamefont{Aseyev}},
  \bibinfo{author}{\bibfnamefont{E.~A.} \bibnamefont{Ryabov}},
  \bibinfo{author}{\bibfnamefont{B.~N.} \bibnamefont{Mironov}},
  \bibnamefont{and} \bibinfo{author}{\bibfnamefont{A.~A.}
  \bibnamefont{Ischenko}}, \bibinfo{journal}{Crystals}
  \textbf{\bibinfo{volume}{10}}, \bibinfo{pages}{452} (\bibinfo{year}{2020}).

\bibitem[{\citenamefont{Baum and Zewail}(2007)}]{BZ07}
\bibinfo{author}{\bibfnamefont{P.}~\bibnamefont{Baum}} \bibnamefont{and}
  \bibinfo{author}{\bibfnamefont{A.~H.} \bibnamefont{Zewail}},
  \bibinfo{journal}{Proc.\ Natl.\ Academ.\ Sci.}
  \textbf{\bibinfo{volume}{104}}, \bibinfo{pages}{18409}
  (\bibinfo{year}{2007}).

\bibitem[{\citenamefont{Sears et~al.}(2008)\citenamefont{Sears, Colby,
  Ischebeck, McGuinness, Nelson, and Noble}}]{SCI08}
\bibinfo{author}{\bibfnamefont{C.~M.~S.} \bibnamefont{Sears}},
  \bibinfo{author}{\bibfnamefont{E.}~\bibnamefont{Colby}},
  \bibinfo{author}{\bibfnamefont{R.}~\bibnamefont{Ischebeck}},
  \bibinfo{author}{\bibfnamefont{C.}~\bibnamefont{McGuinness}},
  \bibinfo{author}{\bibfnamefont{J.}~\bibnamefont{Nelson}}, \bibnamefont{and}
  \bibinfo{author}{\bibfnamefont{R.}~\bibnamefont{Noble}},
  \bibinfo{journal}{Phys.\ Rev.\ Accel.\ Beams} \textbf{\bibinfo{volume}{11}},
  \bibinfo{pages}{061301} (\bibinfo{year}{2008}).

\bibitem[{\citenamefont{Priebe et~al.}(2017)\citenamefont{Priebe, Rathje,
  Yalunin, Hohage, Feist, Sch\"{a}fer, and Ropers}}]{PRY17}
\bibinfo{author}{\bibfnamefont{K.~E.} \bibnamefont{Priebe}},
  \bibinfo{author}{\bibfnamefont{C.}~\bibnamefont{Rathje}},
  \bibinfo{author}{\bibfnamefont{S.~V.} \bibnamefont{Yalunin}},
  \bibinfo{author}{\bibfnamefont{T.}~\bibnamefont{Hohage}},
  \bibinfo{author}{\bibfnamefont{A.}~\bibnamefont{Feist}},
  \bibinfo{author}{\bibfnamefont{S.}~\bibnamefont{Sch\"{a}fer}},
  \bibnamefont{and} \bibinfo{author}{\bibfnamefont{C.}~\bibnamefont{Ropers}},
  \bibinfo{journal}{Nat.\ Photon.} \textbf{\bibinfo{volume}{11}},
  \bibinfo{pages}{793} (\bibinfo{year}{2017}).

\bibitem[{\citenamefont{Morimoto and Baum}(2018{\natexlab{a}})}]{MB18_2}
\bibinfo{author}{\bibfnamefont{Y.}~\bibnamefont{Morimoto}} \bibnamefont{and}
  \bibinfo{author}{\bibfnamefont{P.}~\bibnamefont{Baum}},
  \bibinfo{journal}{Nat.\ Phys.} \textbf{\bibinfo{volume}{14}},
  \bibinfo{pages}{252} (\bibinfo{year}{2018}{\natexlab{a}}).

\bibitem[{\citenamefont{Koz\'ak
  et~al.}(2017{\natexlab{a}})\citenamefont{Koz\'ak, Eckstein, Sch\"onenberger,
  and Hommelhoff}}]{KES17}
\bibinfo{author}{\bibfnamefont{M.}~\bibnamefont{Koz\'ak}},
  \bibinfo{author}{\bibfnamefont{T.}~\bibnamefont{Eckstein}},
  \bibinfo{author}{\bibfnamefont{N.}~\bibnamefont{Sch\"onenberger}},
  \bibnamefont{and}
  \bibinfo{author}{\bibfnamefont{P.}~\bibnamefont{Hommelhoff}},
  \bibinfo{journal}{Nat.\ Phys.} \textbf{\bibinfo{volume}{14}},
  \bibinfo{pages}{121} (\bibinfo{year}{2017}{\natexlab{a}}).

\bibitem[{\citenamefont{Koz\'ak et~al.}(2018)\citenamefont{Koz\'ak,
  Sch\"onenberger, and Hommelhoff}}]{KSH18}
\bibinfo{author}{\bibfnamefont{M.}~\bibnamefont{Koz\'ak}},
  \bibinfo{author}{\bibfnamefont{N.}~\bibnamefont{Sch\"onenberger}},
  \bibnamefont{and}
  \bibinfo{author}{\bibfnamefont{P.}~\bibnamefont{Hommelhoff}},
  \bibinfo{journal}{Phys.\ Rev.\ Lett.} \textbf{\bibinfo{volume}{120}},
  \bibinfo{pages}{103203} (\bibinfo{year}{2018}).

\bibitem[{\citenamefont{Morimoto and Baum}(2018{\natexlab{b}})}]{MB18}
\bibinfo{author}{\bibfnamefont{Y.}~\bibnamefont{Morimoto}} \bibnamefont{and}
  \bibinfo{author}{\bibfnamefont{P.}~\bibnamefont{Baum}},
  \bibinfo{journal}{Phys.\ Rev.\ A} \textbf{\bibinfo{volume}{97}},
  \bibinfo{pages}{033815} (\bibinfo{year}{2018}{\natexlab{b}}).

\bibitem[{\citenamefont{Sch\"onenberger
  et~al.}(2019)\citenamefont{Sch\"onenberger, Mittelbach, Yousefi, McNeur,
  Niedermayer, and Hommelhoff}}]{SMY19}
\bibinfo{author}{\bibfnamefont{N.}~\bibnamefont{Sch\"onenberger}},
  \bibinfo{author}{\bibfnamefont{A.}~\bibnamefont{Mittelbach}},
  \bibinfo{author}{\bibfnamefont{P.}~\bibnamefont{Yousefi}},
  \bibinfo{author}{\bibfnamefont{J.}~\bibnamefont{McNeur}},
  \bibinfo{author}{\bibfnamefont{U.}~\bibnamefont{Niedermayer}},
  \bibnamefont{and}
  \bibinfo{author}{\bibfnamefont{P.}~\bibnamefont{Hommelhoff}},
  \bibinfo{journal}{Phys.\ Rev.\ Lett.} \textbf{\bibinfo{volume}{123}},
  \bibinfo{pages}{264803} (\bibinfo{year}{2019}).

\bibitem[{\citenamefont{Ryabov et~al.}(2020)\citenamefont{Ryabov, Thurner,
  Nabben, Tsarev, and Baum}}]{RTN20}
\bibinfo{author}{\bibfnamefont{A.}~\bibnamefont{Ryabov}},
  \bibinfo{author}{\bibfnamefont{J.~W.} \bibnamefont{Thurner}},
  \bibinfo{author}{\bibfnamefont{D.}~\bibnamefont{Nabben}},
  \bibinfo{author}{\bibfnamefont{M.~V.} \bibnamefont{Tsarev}},
  \bibnamefont{and} \bibinfo{author}{\bibfnamefont{P.}~\bibnamefont{Baum}},
  \bibinfo{journal}{Sci.\ Adv.} \textbf{\bibinfo{volume}{6}},
  \bibinfo{pages}{eabb1393} (\bibinfo{year}{2020}).

\bibitem[{\citenamefont{{Garc\'{\i}a de Abajo}
  et~al.}(2010)\citenamefont{{Garc\'{\i}a de Abajo}, Asenjo-Garcia, and
  Kociak}}]{paper151}
\bibinfo{author}{\bibfnamefont{F.~J.} \bibnamefont{{Garc\'{\i}a de Abajo}}},
  \bibinfo{author}{\bibfnamefont{A.}~\bibnamefont{Asenjo-Garcia}},
  \bibnamefont{and} \bibinfo{author}{\bibfnamefont{M.}~\bibnamefont{Kociak}},
  \bibinfo{journal}{Nano\ Lett.} \textbf{\bibinfo{volume}{10}},
  \bibinfo{pages}{1859} (\bibinfo{year}{2010}).

\bibitem[{\citenamefont{Park et~al.}(2010)\citenamefont{Park, Lin, and
  Zewail}}]{PLZ10}
\bibinfo{author}{\bibfnamefont{S.~T.} \bibnamefont{Park}},
  \bibinfo{author}{\bibfnamefont{M.}~\bibnamefont{Lin}}, \bibnamefont{and}
  \bibinfo{author}{\bibfnamefont{A.~H.} \bibnamefont{Zewail}},
  \bibinfo{journal}{New\ J.\ Phys.} \textbf{\bibinfo{volume}{12}},
  \bibinfo{pages}{123028} (\bibinfo{year}{2010}).

\bibitem[{\citenamefont{Piazza et~al.}(2015)\citenamefont{Piazza, Lummen,
  {Qui\~{n}onez}, Murooka, Reed, Barwick, and Carbone}}]{PLQ15}
\bibinfo{author}{\bibfnamefont{L.}~\bibnamefont{Piazza}},
  \bibinfo{author}{\bibfnamefont{T.~T.~A.} \bibnamefont{Lummen}},
  \bibinfo{author}{\bibfnamefont{E.}~\bibnamefont{{Qui\~{n}onez}}},
  \bibinfo{author}{\bibfnamefont{Y.}~\bibnamefont{Murooka}},
  \bibinfo{author}{\bibfnamefont{B.}~\bibnamefont{Reed}},
  \bibinfo{author}{\bibfnamefont{B.}~\bibnamefont{Barwick}}, \bibnamefont{and}
  \bibinfo{author}{\bibfnamefont{F.}~\bibnamefont{Carbone}},
  \bibinfo{journal}{Nat.\ Commun.} \textbf{\bibinfo{volume}{6}},
  \bibinfo{pages}{6407} (\bibinfo{year}{2015}).

\bibitem[{\citenamefont{Feist et~al.}(2015)\citenamefont{Feist, Echternkamp,
  Schauss, Yalunin, Sch\"afer, and Ropers}}]{FES15}
\bibinfo{author}{\bibfnamefont{A.}~\bibnamefont{Feist}},
  \bibinfo{author}{\bibfnamefont{K.~E.} \bibnamefont{Echternkamp}},
  \bibinfo{author}{\bibfnamefont{J.}~\bibnamefont{Schauss}},
  \bibinfo{author}{\bibfnamefont{S.~V.} \bibnamefont{Yalunin}},
  \bibinfo{author}{\bibfnamefont{S.}~\bibnamefont{Sch\"afer}},
  \bibnamefont{and} \bibinfo{author}{\bibfnamefont{C.}~\bibnamefont{Ropers}},
  \bibinfo{journal}{Nature} \textbf{\bibinfo{volume}{521}},
  \bibinfo{pages}{200} (\bibinfo{year}{2015}).

\bibitem[{\citenamefont{Echternkamp et~al.}(2016)\citenamefont{Echternkamp,
  Feist, Sch\"{a}fer, and Ropers}}]{EFS16}
\bibinfo{author}{\bibfnamefont{K.~E.} \bibnamefont{Echternkamp}},
  \bibinfo{author}{\bibfnamefont{A.}~\bibnamefont{Feist}},
  \bibinfo{author}{\bibfnamefont{S.}~\bibnamefont{Sch\"{a}fer}},
  \bibnamefont{and} \bibinfo{author}{\bibfnamefont{C.}~\bibnamefont{Ropers}},
  \bibinfo{journal}{Nat.\ Phys.} \textbf{\bibinfo{volume}{12}},
  \bibinfo{pages}{1000} (\bibinfo{year}{2016}).

\bibitem[{\citenamefont{Vanacore et~al.}(2018)\citenamefont{Vanacore, Madan,
  Berruto, Wang, Pomarico, Lamb, McGrouther, Kaminer, Barwick, {Garc\'{\i}a de
  Abajo} et~al.}}]{paper311}
\bibinfo{author}{\bibfnamefont{G.~M.} \bibnamefont{Vanacore}},
  \bibinfo{author}{\bibfnamefont{I.}~\bibnamefont{Madan}},
  \bibinfo{author}{\bibfnamefont{G.}~\bibnamefont{Berruto}},
  \bibinfo{author}{\bibfnamefont{K.}~\bibnamefont{Wang}},
  \bibinfo{author}{\bibfnamefont{E.}~\bibnamefont{Pomarico}},
  \bibinfo{author}{\bibfnamefont{R.~J.} \bibnamefont{Lamb}},
  \bibinfo{author}{\bibfnamefont{D.}~\bibnamefont{McGrouther}},
  \bibinfo{author}{\bibfnamefont{I.}~\bibnamefont{Kaminer}},
  \bibinfo{author}{\bibfnamefont{B.}~\bibnamefont{Barwick}},
  \bibinfo{author}{\bibfnamefont{F.~J.} \bibnamefont{{Garc\'{\i}a de Abajo}}},
  \bibnamefont{et~al.}, \bibinfo{journal}{Nat.\ Commun.}
  \textbf{\bibinfo{volume}{9}}, \bibinfo{pages}{2694} (\bibinfo{year}{2018}).

\bibitem[{\citenamefont{Cai et~al.}(2018)\citenamefont{Cai, Reinhardt, Kaminer,
  and {Garc\'{\i}a de Abajo}}}]{paper312}
\bibinfo{author}{\bibfnamefont{W.}~\bibnamefont{Cai}},
  \bibinfo{author}{\bibfnamefont{O.}~\bibnamefont{Reinhardt}},
  \bibinfo{author}{\bibfnamefont{I.}~\bibnamefont{Kaminer}}, \bibnamefont{and}
  \bibinfo{author}{\bibfnamefont{F.~J.} \bibnamefont{{Garc\'{\i}a de Abajo}}},
  \bibinfo{journal}{Phys.\ Rev.\ B} \textbf{\bibinfo{volume}{98}},
  \bibinfo{pages}{045424} (\bibinfo{year}{2018}).

\bibitem[{\citenamefont{Vanacore et~al.}(2019)\citenamefont{Vanacore, Berruto,
  Madan, Pomarico, Biagioni, Lamb, McGrouther, Reinhardt, Kaminer, Barwick
  et~al.}}]{paper332}
\bibinfo{author}{\bibfnamefont{G.~M.} \bibnamefont{Vanacore}},
  \bibinfo{author}{\bibfnamefont{G.}~\bibnamefont{Berruto}},
  \bibinfo{author}{\bibfnamefont{I.}~\bibnamefont{Madan}},
  \bibinfo{author}{\bibfnamefont{E.}~\bibnamefont{Pomarico}},
  \bibinfo{author}{\bibfnamefont{P.}~\bibnamefont{Biagioni}},
  \bibinfo{author}{\bibfnamefont{R.~J.} \bibnamefont{Lamb}},
  \bibinfo{author}{\bibfnamefont{D.}~\bibnamefont{McGrouther}},
  \bibinfo{author}{\bibfnamefont{O.}~\bibnamefont{Reinhardt}},
  \bibinfo{author}{\bibfnamefont{I.}~\bibnamefont{Kaminer}},
  \bibinfo{author}{\bibfnamefont{B.}~\bibnamefont{Barwick}},
  \bibnamefont{et~al.}, \bibinfo{journal}{Nat.\ Mater.}
  \textbf{\bibinfo{volume}{18}}, \bibinfo{pages}{573} (\bibinfo{year}{2019}).

\bibitem[{\citenamefont{Feist et~al.}(2020)\citenamefont{Feist, Yalunin,
  Sch\"afer, and Ropers}}]{FYS20}
\bibinfo{author}{\bibfnamefont{A.}~\bibnamefont{Feist}},
  \bibinfo{author}{\bibfnamefont{S.~V.} \bibnamefont{Yalunin}},
  \bibinfo{author}{\bibfnamefont{S.}~\bibnamefont{Sch\"afer}},
  \bibnamefont{and} \bibinfo{author}{\bibfnamefont{C.}~\bibnamefont{Ropers}},
  \bibinfo{journal}{Phys.\ Rev.\ Research} \textbf{\bibinfo{volume}{2}},
  \bibinfo{pages}{043227} (\bibinfo{year}{2020}).

\bibitem[{\citenamefont{Park and Zewail}(2012)}]{PZ12}
\bibinfo{author}{\bibfnamefont{S.~T.} \bibnamefont{Park}} \bibnamefont{and}
  \bibinfo{author}{\bibfnamefont{A.~H.} \bibnamefont{Zewail}},
  \bibinfo{journal}{J.\ Phys.\ Chem.\ A} \textbf{\bibinfo{volume}{116}},
  \bibinfo{pages}{11128} (\bibinfo{year}{2012}).

\bibitem[{\citenamefont{Kirchner et~al.}(2014)\citenamefont{Kirchner, Gliserin,
  Krausz, and Baum}}]{KGK14}
\bibinfo{author}{\bibfnamefont{F.~O.} \bibnamefont{Kirchner}},
  \bibinfo{author}{\bibfnamefont{A.}~\bibnamefont{Gliserin}},
  \bibinfo{author}{\bibfnamefont{F.}~\bibnamefont{Krausz}}, \bibnamefont{and}
  \bibinfo{author}{\bibfnamefont{P.}~\bibnamefont{Baum}},
  \bibinfo{journal}{Nat.\ Photon.} \textbf{\bibinfo{volume}{8}},
  \bibinfo{pages}{52} (\bibinfo{year}{2014}).

\bibitem[{\citenamefont{Lummen et~al.}(2016)\citenamefont{Lummen, Lamb,
  Berruto, LaGrange, Negro, {Garc\'{\i}a de Abajo}, McGrouther, Barwick, and
  Carbone}}]{paper282}
\bibinfo{author}{\bibfnamefont{T.~T.~A.} \bibnamefont{Lummen}},
  \bibinfo{author}{\bibfnamefont{R.~J.} \bibnamefont{Lamb}},
  \bibinfo{author}{\bibfnamefont{G.}~\bibnamefont{Berruto}},
  \bibinfo{author}{\bibfnamefont{T.}~\bibnamefont{LaGrange}},
  \bibinfo{author}{\bibfnamefont{L.~D.} \bibnamefont{Negro}},
  \bibinfo{author}{\bibfnamefont{F.~J.} \bibnamefont{{Garc\'{\i}a de Abajo}}},
  \bibinfo{author}{\bibfnamefont{D.}~\bibnamefont{McGrouther}},
  \bibinfo{author}{\bibfnamefont{B.}~\bibnamefont{Barwick}}, \bibnamefont{and}
  \bibinfo{author}{\bibfnamefont{F.}~\bibnamefont{Carbone}},
  \bibinfo{journal}{Nat.\ Commun.} \textbf{\bibinfo{volume}{7}},
  \bibinfo{pages}{13156} (\bibinfo{year}{2016}).

\bibitem[{\citenamefont{Kealhofer et~al.}(2016)\citenamefont{Kealhofer,
  Schneider, Ehberger, Ryabov, Krausz, and Baum}}]{KSE16}
\bibinfo{author}{\bibfnamefont{C.}~\bibnamefont{Kealhofer}},
  \bibinfo{author}{\bibfnamefont{W.}~\bibnamefont{Schneider}},
  \bibinfo{author}{\bibfnamefont{D.}~\bibnamefont{Ehberger}},
  \bibinfo{author}{\bibfnamefont{A.}~\bibnamefont{Ryabov}},
  \bibinfo{author}{\bibfnamefont{F.}~\bibnamefont{Krausz}}, \bibnamefont{and}
  \bibinfo{author}{\bibfnamefont{P.}~\bibnamefont{Baum}},
  \bibinfo{journal}{Science} \textbf{\bibinfo{volume}{352}},
  \bibinfo{pages}{429} (\bibinfo{year}{2016}).

\bibitem[{\citenamefont{Ryabov and Baum}(2016)}]{RB16}
\bibinfo{author}{\bibfnamefont{A.}~\bibnamefont{Ryabov}} \bibnamefont{and}
  \bibinfo{author}{\bibfnamefont{P.}~\bibnamefont{Baum}},
  \bibinfo{journal}{Science} \textbf{\bibinfo{volume}{353}},
  \bibinfo{pages}{374} (\bibinfo{year}{2016}).

\bibitem[{\citenamefont{Vanacore et~al.}(2016)\citenamefont{Vanacore,
  Fitzpatrick, and Zewail}}]{VFZ16}
\bibinfo{author}{\bibfnamefont{G.~M.} \bibnamefont{Vanacore}},
  \bibinfo{author}{\bibfnamefont{A.~W.~P.} \bibnamefont{Fitzpatrick}},
  \bibnamefont{and} \bibinfo{author}{\bibfnamefont{A.~H.}
  \bibnamefont{Zewail}}, \bibinfo{journal}{Nano\ Today}
  \textbf{\bibinfo{volume}{11}}, \bibinfo{pages}{228} (\bibinfo{year}{2016}).

\bibitem[{\citenamefont{{Garc\'{\i}a de Abajo}
  et~al.}(2016)\citenamefont{{Garc\'{\i}a de Abajo}, Barwick, and
  Carbone}}]{paper272}
\bibinfo{author}{\bibfnamefont{F.~J.} \bibnamefont{{Garc\'{\i}a de Abajo}}},
  \bibinfo{author}{\bibfnamefont{B.}~\bibnamefont{Barwick}}, \bibnamefont{and}
  \bibinfo{author}{\bibfnamefont{F.}~\bibnamefont{Carbone}},
  \bibinfo{journal}{Phys.\ Rev.\ B} \textbf{\bibinfo{volume}{94}},
  \bibinfo{pages}{041404(R)} (\bibinfo{year}{2016}).

\bibitem[{\citenamefont{Koz\'ak
  et~al.}(2017{\natexlab{b}})\citenamefont{Koz\'ak, McNeur, Leedle, Deng,
  Sch\"onenberger, Ruehl, Hartl, Harris, Byer, and Hommelhoff}}]{KML17}
\bibinfo{author}{\bibfnamefont{M.}~\bibnamefont{Koz\'ak}},
  \bibinfo{author}{\bibfnamefont{J.}~\bibnamefont{McNeur}},
  \bibinfo{author}{\bibfnamefont{K.~J.} \bibnamefont{Leedle}},
  \bibinfo{author}{\bibfnamefont{H.}~\bibnamefont{Deng}},
  \bibinfo{author}{\bibfnamefont{N.}~\bibnamefont{Sch\"onenberger}},
  \bibinfo{author}{\bibfnamefont{A.}~\bibnamefont{Ruehl}},
  \bibinfo{author}{\bibfnamefont{I.}~\bibnamefont{Hartl}},
  \bibinfo{author}{\bibfnamefont{J.~S.} \bibnamefont{Harris}},
  \bibinfo{author}{\bibfnamefont{R.~L.} \bibnamefont{Byer}}, \bibnamefont{and}
  \bibinfo{author}{\bibfnamefont{P.}~\bibnamefont{Hommelhoff}},
  \bibinfo{journal}{Nat.\ Commun.} \textbf{\bibinfo{volume}{8}},
  \bibinfo{pages}{14342} (\bibinfo{year}{2017}{\natexlab{b}}).

\bibitem[{\citenamefont{Feist et~al.}(2017)\citenamefont{Feist, Bach,
  N.~Rubiano~{da Silva}, M\"{a}ller, Priebe, Domr\"{a}se, Gatzmann, Rost,
  Schauss, Strauch et~al.}}]{FBR17}
\bibinfo{author}{\bibfnamefont{A.}~\bibnamefont{Feist}},
  \bibinfo{author}{\bibfnamefont{N.}~\bibnamefont{Bach}},
  \bibinfo{author}{\bibfnamefont{T.~D.} \bibnamefont{N.~Rubiano~{da Silva}}},
  \bibinfo{author}{\bibfnamefont{M.}~\bibnamefont{M\"{a}ller}},
  \bibinfo{author}{\bibfnamefont{K.~E.} \bibnamefont{Priebe}},
  \bibinfo{author}{\bibfnamefont{T.}~\bibnamefont{Domr\"{a}se}},
  \bibinfo{author}{\bibfnamefont{J.~G.} \bibnamefont{Gatzmann}},
  \bibinfo{author}{\bibfnamefont{S.}~\bibnamefont{Rost}},
  \bibinfo{author}{\bibfnamefont{J.}~\bibnamefont{Schauss}},
  \bibinfo{author}{\bibfnamefont{S.}~\bibnamefont{Strauch}},
  \bibnamefont{et~al.}, \bibinfo{journal}{Ultramicroscopy}
  \textbf{\bibinfo{volume}{176}}, \bibinfo{pages}{63} (\bibinfo{year}{2017}).

\bibitem[{\citenamefont{Pomarico et~al.}(2018)\citenamefont{Pomarico, Madan,
  Berruto, Vanacore, Wang, Kaminer, {Garc\'{\i}a de Abajo}, and
  Carbone}}]{paper306}
\bibinfo{author}{\bibfnamefont{E.}~\bibnamefont{Pomarico}},
  \bibinfo{author}{\bibfnamefont{I.}~\bibnamefont{Madan}},
  \bibinfo{author}{\bibfnamefont{G.}~\bibnamefont{Berruto}},
  \bibinfo{author}{\bibfnamefont{G.~M.} \bibnamefont{Vanacore}},
  \bibinfo{author}{\bibfnamefont{K.}~\bibnamefont{Wang}},
  \bibinfo{author}{\bibfnamefont{I.}~\bibnamefont{Kaminer}},
  \bibinfo{author}{\bibfnamefont{F.~J.} \bibnamefont{{Garc\'{\i}a de Abajo}}},
  \bibnamefont{and} \bibinfo{author}{\bibfnamefont{F.}~\bibnamefont{Carbone}},
  \bibinfo{journal}{ACS\ Photonics} \textbf{\bibinfo{volume}{5}},
  \bibinfo{pages}{759} (\bibinfo{year}{2018}).

\bibitem[{\citenamefont{Das et~al.}(2019)\citenamefont{Das, Blazit, Tenc\'e,
  Zagonel, Auad, Lee, Ling, Losquin, C.~Colliex, {Garc\'{\i}a de Abajo}
  et~al.}}]{paper325}
\bibinfo{author}{\bibfnamefont{P.}~\bibnamefont{Das}},
  \bibinfo{author}{\bibfnamefont{J.~D.} \bibnamefont{Blazit}},
  \bibinfo{author}{\bibfnamefont{M.}~\bibnamefont{Tenc\'e}},
  \bibinfo{author}{\bibfnamefont{L.~F.} \bibnamefont{Zagonel}},
  \bibinfo{author}{\bibfnamefont{Y.}~\bibnamefont{Auad}},
  \bibinfo{author}{\bibfnamefont{Y.~H.} \bibnamefont{Lee}},
  \bibinfo{author}{\bibfnamefont{X.~Y.} \bibnamefont{Ling}},
  \bibinfo{author}{\bibfnamefont{A.}~\bibnamefont{Losquin}},
  \bibinfo{author}{\bibfnamefont{O.~S.} \bibnamefont{C.~Colliex}},
  \bibinfo{author}{\bibfnamefont{F.~J.} \bibnamefont{{Garc\'{\i}a de Abajo}}},
  \bibnamefont{et~al.}, \bibinfo{journal}{Ultramicroscopy}
  \textbf{\bibinfo{volume}{203}}, \bibinfo{pages}{44} (\bibinfo{year}{2019}).

\bibitem[{\citenamefont{Kfir}(2019)}]{K19}
\bibinfo{author}{\bibfnamefont{O.}~\bibnamefont{Kfir}},
  \bibinfo{journal}{Phys.\ Rev.\ Lett.} \textbf{\bibinfo{volume}{123}},
  \bibinfo{pages}{103602} (\bibinfo{year}{2019}).

\bibitem[{\citenamefont{Pan et~al.}(2019)\citenamefont{Pan, Zhang, and
  Gover}}]{PZG19}
\bibinfo{author}{\bibfnamefont{Y.}~\bibnamefont{Pan}},
  \bibinfo{author}{\bibfnamefont{B.}~\bibnamefont{Zhang}}, \bibnamefont{and}
  \bibinfo{author}{\bibfnamefont{A.}~\bibnamefont{Gover}},
  \bibinfo{journal}{Phys.\ Rev.\ Lett.} \textbf{\bibinfo{volume}{122}},
  \bibinfo{pages}{183204} (\bibinfo{year}{2019}).

\bibitem[{\citenamefont{{Di Giulio} et~al.}(2019)\citenamefont{{Di Giulio},
  Kociak, and {Garc\'{\i}a de Abajo}}}]{paper339}
\bibinfo{author}{\bibfnamefont{V.}~\bibnamefont{{Di Giulio}}},
  \bibinfo{author}{\bibfnamefont{M.}~\bibnamefont{Kociak}}, \bibnamefont{and}
  \bibinfo{author}{\bibfnamefont{F.~J.} \bibnamefont{{Garc\'{\i}a de Abajo}}},
  \bibinfo{journal}{Optica} \textbf{\bibinfo{volume}{6}}, \bibinfo{pages}{1524}
  (\bibinfo{year}{2019}).

\bibitem[{\citenamefont{Reinhardt et~al.}(2020)\citenamefont{Reinhardt, Mechel,
  Lynch, and Kaminer}}]{RML20}
\bibinfo{author}{\bibfnamefont{O.}~\bibnamefont{Reinhardt}},
  \bibinfo{author}{\bibfnamefont{C.}~\bibnamefont{Mechel}},
  \bibinfo{author}{\bibfnamefont{M.}~\bibnamefont{Lynch}}, \bibnamefont{and}
  \bibinfo{author}{\bibfnamefont{I.}~\bibnamefont{Kaminer}},
  \bibinfo{journal}{Ann.\ Phys.} \textbf{\bibinfo{volume}{533}},
  \bibinfo{pages}{2000254} (\bibinfo{year}{2020}).

\bibitem[{\citenamefont{Dahan et~al.}(2020)\citenamefont{Dahan, Nehemia,
  Shentcis, Reinhardt, Adiv, Shi, Be'er, Lynch, Kurman, Wang et~al.}}]{DNS20}
\bibinfo{author}{\bibfnamefont{R.}~\bibnamefont{Dahan}},
  \bibinfo{author}{\bibfnamefont{S.}~\bibnamefont{Nehemia}},
  \bibinfo{author}{\bibfnamefont{M.}~\bibnamefont{Shentcis}},
  \bibinfo{author}{\bibfnamefont{O.}~\bibnamefont{Reinhardt}},
  \bibinfo{author}{\bibfnamefont{Y.}~\bibnamefont{Adiv}},
  \bibinfo{author}{\bibfnamefont{X.}~\bibnamefont{Shi}},
  \bibinfo{author}{\bibfnamefont{O.}~\bibnamefont{Be'er}},
  \bibinfo{author}{\bibfnamefont{M.~H.} \bibnamefont{Lynch}},
  \bibinfo{author}{\bibfnamefont{Y.}~\bibnamefont{Kurman}},
  \bibinfo{author}{\bibfnamefont{K.}~\bibnamefont{Wang}}, \bibnamefont{et~al.},
  \bibinfo{journal}{Nat.\ Phys.} \textbf{\bibinfo{volume}{16}},
  \bibinfo{pages}{1123} (\bibinfo{year}{2020}).

\bibitem[{\citenamefont{Kfir et~al.}(2020)\citenamefont{Kfir,
  Louren\c{c}o-Martins, Storeck, Sivis, Harvey, Kippenberg, Feist, and
  Ropers}}]{KLS20}
\bibinfo{author}{\bibfnamefont{O.}~\bibnamefont{Kfir}},
  \bibinfo{author}{\bibfnamefont{H.}~\bibnamefont{Louren\c{c}o-Martins}},
  \bibinfo{author}{\bibfnamefont{G.}~\bibnamefont{Storeck}},
  \bibinfo{author}{\bibfnamefont{M.}~\bibnamefont{Sivis}},
  \bibinfo{author}{\bibfnamefont{T.~R.} \bibnamefont{Harvey}},
  \bibinfo{author}{\bibfnamefont{T.~J.} \bibnamefont{Kippenberg}},
  \bibinfo{author}{\bibfnamefont{A.}~\bibnamefont{Feist}}, \bibnamefont{and}
  \bibinfo{author}{\bibfnamefont{C.}~\bibnamefont{Ropers}},
  \bibinfo{journal}{Nature} \textbf{\bibinfo{volume}{582}}, \bibinfo{pages}{46}
  (\bibinfo{year}{2020}).

\bibitem[{\citenamefont{Wang et~al.}(2020)\citenamefont{Wang, Dahan, Shentcis,
  Kauffmann, Hayun, Reinhardt, Tsesses, and Kaminer}}]{WDS20}
\bibinfo{author}{\bibfnamefont{K.}~\bibnamefont{Wang}},
  \bibinfo{author}{\bibfnamefont{R.}~\bibnamefont{Dahan}},
  \bibinfo{author}{\bibfnamefont{M.}~\bibnamefont{Shentcis}},
  \bibinfo{author}{\bibfnamefont{Y.}~\bibnamefont{Kauffmann}},
  \bibinfo{author}{\bibfnamefont{A.~B.} \bibnamefont{Hayun}},
  \bibinfo{author}{\bibfnamefont{O.}~\bibnamefont{Reinhardt}},
  \bibinfo{author}{\bibfnamefont{S.}~\bibnamefont{Tsesses}}, \bibnamefont{and}
  \bibinfo{author}{\bibfnamefont{I.}~\bibnamefont{Kaminer}},
  \bibinfo{journal}{Nature} \textbf{\bibinfo{volume}{582}}, \bibinfo{pages}{50}
  (\bibinfo{year}{2020}).

\bibitem[{\citenamefont{Reinhardt and Kaminer}(2020)}]{RK20}
\bibinfo{author}{\bibfnamefont{O.}~\bibnamefont{Reinhardt}} \bibnamefont{and}
  \bibinfo{author}{\bibfnamefont{I.}~\bibnamefont{Kaminer}},
  \bibinfo{journal}{ACS\ Photonics} \textbf{\bibinfo{volume}{7}},
  \bibinfo{pages}{2859} (\bibinfo{year}{2020}).

\bibitem[{\citenamefont{Madan et~al.}(2020)\citenamefont{Madan, Vanacore,
  Gargiulo, LaGrange, and Carbone}}]{MVG20}
\bibinfo{author}{\bibfnamefont{I.}~\bibnamefont{Madan}},
  \bibinfo{author}{\bibfnamefont{G.~M.} \bibnamefont{Vanacore}},
  \bibinfo{author}{\bibfnamefont{S.}~\bibnamefont{Gargiulo}},
  \bibinfo{author}{\bibfnamefont{T.}~\bibnamefont{LaGrange}}, \bibnamefont{and}
  \bibinfo{author}{\bibfnamefont{F.}~\bibnamefont{Carbone}},
  \bibinfo{journal}{Appl.\ Phys.\ Lett.} \textbf{\bibinfo{volume}{116}},
  \bibinfo{pages}{230502} (\bibinfo{year}{2020}).

\bibitem[{\citenamefont{{Di Giulio} and {Garc\'{\i}a de
  Abajo}}(2020)}]{paper360}
\bibinfo{author}{\bibfnamefont{V.}~\bibnamefont{{Di Giulio}}} \bibnamefont{and}
  \bibinfo{author}{\bibfnamefont{F.~J.} \bibnamefont{{Garc\'{\i}a de Abajo}}},
  \bibinfo{journal}{Optica} \textbf{\bibinfo{volume}{7}}, \bibinfo{pages}{1820}
  (\bibinfo{year}{2020}).

\bibitem[{\citenamefont{Vanacore et~al.}(2020)\citenamefont{Vanacore, Madan,
  and Carbone}}]{VMC20}
\bibinfo{author}{\bibfnamefont{G.~M.} \bibnamefont{Vanacore}},
  \bibinfo{author}{\bibfnamefont{I.}~\bibnamefont{Madan}}, \bibnamefont{and}
  \bibinfo{author}{\bibfnamefont{F.}~\bibnamefont{Carbone}},
  \bibinfo{journal}{Riv.\ Nuovo\ Cimento} \textbf{\bibinfo{volume}{43}},
  \bibinfo{pages}{567} (\bibinfo{year}{2020}).

\bibitem[{\citenamefont{Kurman et~al.}(2021)\citenamefont{Kurman, Dahan,
  Sheinfux, Wang, Yannai, Adiv, Reinhardt, Tizei, Woo, Li et~al.}}]{KDS21}
\bibinfo{author}{\bibfnamefont{Y.}~\bibnamefont{Kurman}},
  \bibinfo{author}{\bibfnamefont{R.}~\bibnamefont{Dahan}},
  \bibinfo{author}{\bibfnamefont{H.~H.} \bibnamefont{Sheinfux}},
  \bibinfo{author}{\bibfnamefont{K.}~\bibnamefont{Wang}},
  \bibinfo{author}{\bibfnamefont{M.}~\bibnamefont{Yannai}},
  \bibinfo{author}{\bibfnamefont{Y.}~\bibnamefont{Adiv}},
  \bibinfo{author}{\bibfnamefont{O.}~\bibnamefont{Reinhardt}},
  \bibinfo{author}{\bibfnamefont{L.~H.~G.} \bibnamefont{Tizei}},
  \bibinfo{author}{\bibfnamefont{S.~Y.} \bibnamefont{Woo}},
  \bibinfo{author}{\bibfnamefont{J.}~\bibnamefont{Li}}, \bibnamefont{et~al.},
  \bibinfo{journal}{Science} \textbf{\bibinfo{volume}{372}},
  \bibinfo{pages}{1181} (\bibinfo{year}{2021}).

\bibitem[{\citenamefont{Henke et~al.}(2021)\citenamefont{Henke, Raja, Feist,
  Huang, Arend, Y.~Yang, Wang, M\"oller, Pan, J.~Liu et~al.}}]{HRF21}
\bibinfo{author}{\bibfnamefont{J.-W.} \bibnamefont{Henke}},
  \bibinfo{author}{\bibfnamefont{A.~S.} \bibnamefont{Raja}},
  \bibinfo{author}{\bibfnamefont{A.}~\bibnamefont{Feist}},
  \bibinfo{author}{\bibfnamefont{G.}~\bibnamefont{Huang}},
  \bibinfo{author}{\bibfnamefont{G.}~\bibnamefont{Arend}},
  \bibinfo{author}{\bibfnamefont{J.~K.} \bibnamefont{Y.~Yang}},
  \bibinfo{author}{\bibfnamefont{R.~N.} \bibnamefont{Wang}},
  \bibinfo{author}{\bibfnamefont{M.}~\bibnamefont{M\"oller}},
  \bibinfo{author}{\bibfnamefont{J.}~\bibnamefont{Pan}},
  \bibinfo{author}{\bibfnamefont{O.~K.} \bibnamefont{J.~Liu}},
  \bibnamefont{et~al.}, \emph{\bibinfo{title}{Integrated photonics enables
  continuous-beam electron phase modulation}} (\bibinfo{year}{2021}),
  \eprint{2105.03729}.

\bibitem[{\citenamefont{Kone\v{c}n\'{a}
  et~al.}(2019)\citenamefont{Kone\v{c}n\'{a}, {Di Giulio}, Mkhitaryan, Ropers,
  and {Garc\'{\i}a de Abajo}}}]{paper347}
\bibinfo{author}{\bibfnamefont{A.}~\bibnamefont{Kone\v{c}n\'{a}}},
  \bibinfo{author}{\bibfnamefont{V.}~\bibnamefont{{Di Giulio}}},
  \bibinfo{author}{\bibfnamefont{V.}~\bibnamefont{Mkhitaryan}},
  \bibinfo{author}{\bibfnamefont{C.}~\bibnamefont{Ropers}}, \bibnamefont{and}
  \bibinfo{author}{\bibfnamefont{F.~J.} \bibnamefont{{Garc\'{\i}a de Abajo}}},
  \bibinfo{journal}{ACS\ Photonics} \textbf{\bibinfo{volume}{7}},
  \bibinfo{pages}{1290} (\bibinfo{year}{2019}).

\bibitem[{\citenamefont{Shirley}(1965)}]{S1965}
\bibinfo{author}{\bibfnamefont{J.~H.} \bibnamefont{Shirley}},
  \bibinfo{journal}{Phys.\ Rev.} \textbf{\bibinfo{volume}{138}},
  \bibinfo{pages}{B979} (\bibinfo{year}{1965}).

\bibitem[{\citenamefont{Lindner et~al.}(2011)\citenamefont{Lindner, Refael, and
  Galitski}}]{LRG11}
\bibinfo{author}{\bibfnamefont{N.~H.} \bibnamefont{Lindner}},
  \bibinfo{author}{\bibfnamefont{G.}~\bibnamefont{Refael}}, \bibnamefont{and}
  \bibinfo{author}{\bibfnamefont{V.}~\bibnamefont{Galitski}},
  \bibinfo{journal}{Nat.\ Phys.} \textbf{\bibinfo{volume}{7}},
  \bibinfo{pages}{490} (\bibinfo{year}{2011}).

\bibitem[{\citenamefont{Sie et~al.}(2015)\citenamefont{Sie, McIver, Lee, Fu,
  Kong, and Gedik}}]{SML15}
\bibinfo{author}{\bibfnamefont{E.~J.} \bibnamefont{Sie}},
  \bibinfo{author}{\bibfnamefont{J.~W.} \bibnamefont{McIver}},
  \bibinfo{author}{\bibfnamefont{Y.-H.} \bibnamefont{Lee}},
  \bibinfo{author}{\bibfnamefont{L.}~\bibnamefont{Fu}},
  \bibinfo{author}{\bibfnamefont{J.}~\bibnamefont{Kong}}, \bibnamefont{and}
  \bibinfo{author}{\bibfnamefont{N.}~\bibnamefont{Gedik}},
  \bibinfo{journal}{Nat.\ Mater.} \textbf{\bibinfo{volume}{14}},
  \bibinfo{pages}{290} (\bibinfo{year}{2015}).

\bibitem[{\citenamefont{Jotzu et~al.}(2014)\citenamefont{Jotzu, Messer,
  Desbuquois, Lebrat, Uehlinger, Greif, and Esslinger}}]{JMD14}
\bibinfo{author}{\bibfnamefont{G.}~\bibnamefont{Jotzu}},
  \bibinfo{author}{\bibfnamefont{M.}~\bibnamefont{Messer}},
  \bibinfo{author}{\bibfnamefont{R.}~\bibnamefont{Desbuquois}},
  \bibinfo{author}{\bibfnamefont{M.}~\bibnamefont{Lebrat}},
  \bibinfo{author}{\bibfnamefont{T.}~\bibnamefont{Uehlinger}},
  \bibinfo{author}{\bibfnamefont{D.}~\bibnamefont{Greif}}, \bibnamefont{and}
  \bibinfo{author}{\bibfnamefont{T.}~\bibnamefont{Esslinger}},
  \bibinfo{journal}{Nature} \textbf{\bibinfo{volume}{515}},
  \bibinfo{pages}{237} (\bibinfo{year}{2014}).

\bibitem[{\citenamefont{Rechtsman et~al.}(2013)\citenamefont{Rechtsman, Zeuner,
  Plotnik, Lumer, Podolsky, Dreisow, Nolte, Segev, and Szameit}}]{RZP13}
\bibinfo{author}{\bibfnamefont{M.~C.} \bibnamefont{Rechtsman}},
  \bibinfo{author}{\bibfnamefont{J.~M.} \bibnamefont{Zeuner}},
  \bibinfo{author}{\bibfnamefont{Y.}~\bibnamefont{Plotnik}},
  \bibinfo{author}{\bibfnamefont{Y.}~\bibnamefont{Lumer}},
  \bibinfo{author}{\bibfnamefont{D.}~\bibnamefont{Podolsky}},
  \bibinfo{author}{\bibfnamefont{F.}~\bibnamefont{Dreisow}},
  \bibinfo{author}{\bibfnamefont{S.}~\bibnamefont{Nolte}},
  \bibinfo{author}{\bibfnamefont{M.}~\bibnamefont{Segev}}, \bibnamefont{and}
  \bibinfo{author}{\bibfnamefont{A.}~\bibnamefont{Szameit}},
  \bibinfo{journal}{Nature} \textbf{\bibinfo{volume}{496}},
  \bibinfo{pages}{196} (\bibinfo{year}{2013}).

\bibitem[{\citenamefont{Beatrez et~al.}(2021)\citenamefont{Beatrez, Janes,
  Akkiraju, Pillai, Oddo, Reshetikhin, Druga, McAllister, Elo, Gilbert
  et~al.}}]{BJA21}
\bibinfo{author}{\bibfnamefont{W.}~\bibnamefont{Beatrez}},
  \bibinfo{author}{\bibfnamefont{O.}~\bibnamefont{Janes}},
  \bibinfo{author}{\bibfnamefont{A.}~\bibnamefont{Akkiraju}},
  \bibinfo{author}{\bibfnamefont{A.}~\bibnamefont{Pillai}},
  \bibinfo{author}{\bibfnamefont{A.}~\bibnamefont{Oddo}},
  \bibinfo{author}{\bibfnamefont{P.}~\bibnamefont{Reshetikhin}},
  \bibinfo{author}{\bibfnamefont{E.}~\bibnamefont{Druga}},
  \bibinfo{author}{\bibfnamefont{M.}~\bibnamefont{McAllister}},
  \bibinfo{author}{\bibfnamefont{M.}~\bibnamefont{Elo}},
  \bibinfo{author}{\bibfnamefont{B.}~\bibnamefont{Gilbert}},
  \bibnamefont{et~al.}, \bibinfo{journal}{Phys.\ Rev.\ Lett.}
  \textbf{\bibinfo{volume}{127}}, \bibinfo{pages}{170603}
  (\bibinfo{year}{2021}).

\bibitem[{\citenamefont{Gover and Yariv}(2020)}]{GY20}
\bibinfo{author}{\bibfnamefont{A.}~\bibnamefont{Gover}} \bibnamefont{and}
  \bibinfo{author}{\bibfnamefont{A.}~\bibnamefont{Yariv}},
  \bibinfo{journal}{Phys.\ Rev.\ Lett.} \textbf{\bibinfo{volume}{124}},
  \bibinfo{pages}{064801} (\bibinfo{year}{2020}).

\bibitem[{\citenamefont{Zhao et~al.}(2021)\citenamefont{Zhao, Sun, and
  Fan}}]{ZSF21}
\bibinfo{author}{\bibfnamefont{Z.}~\bibnamefont{Zhao}},
  \bibinfo{author}{\bibfnamefont{X.-Q.} \bibnamefont{Sun}}, \bibnamefont{and}
  \bibinfo{author}{\bibfnamefont{S.}~\bibnamefont{Fan}},
  \emph{\bibinfo{title}{Quantum entanglement and modulation enhancement of
  free-electron--bound-electron interaction}} (\bibinfo{year}{2021}).

\bibitem[{\citenamefont{{Garc\'{\i}a de Abajo} and {Di
  Giulio}}(2020)}]{paper371}
\bibinfo{author}{\bibfnamefont{F.~J.} \bibnamefont{{Garc\'{\i}a de Abajo}}}
  \bibnamefont{and} \bibinfo{author}{\bibfnamefont{V.}~\bibnamefont{{Di
  Giulio}}}, \bibinfo{journal}{ACS\ Photonics} \textbf{\bibinfo{volume}{8}},
  \bibinfo{pages}{945} (\bibinfo{year}{2020}).

\bibitem[{\citenamefont{Kfir et~al.}(2021)\citenamefont{Kfir, {Di Giulio},
  {Garc\'{\i}a de Abajo}, and Ropers}}]{paper374}
\bibinfo{author}{\bibfnamefont{O.}~\bibnamefont{Kfir}},
  \bibinfo{author}{\bibfnamefont{V.}~\bibnamefont{{Di Giulio}}},
  \bibinfo{author}{\bibfnamefont{F.~J.} \bibnamefont{{Garc\'{\i}a de Abajo}}},
  \bibnamefont{and} \bibinfo{author}{\bibfnamefont{C.}~\bibnamefont{Ropers}},
  \bibinfo{journal}{Sci.\ Adv.} \textbf{\bibinfo{volume}{7}},
  \bibinfo{pages}{eabf6380} (\bibinfo{year}{2021}).

\bibitem[{\citenamefont{{Di Giulio} et~al.}(2021)\citenamefont{{Di Giulio},
  Kfir, Ropers, and {Garc\'{\i}a de Abajo}}}]{paper373}
\bibinfo{author}{\bibfnamefont{V.}~\bibnamefont{{Di Giulio}}},
  \bibinfo{author}{\bibfnamefont{O.}~\bibnamefont{Kfir}},
  \bibinfo{author}{\bibfnamefont{C.}~\bibnamefont{Ropers}}, \bibnamefont{and}
  \bibinfo{author}{\bibfnamefont{F.~J.} \bibnamefont{{Garc\'{\i}a de Abajo}}},
  \bibinfo{journal}{ACS\ Nano} \textbf{\bibinfo{volume}{15}},
  \bibinfo{pages}{7290} (\bibinfo{year}{2021}).

\bibitem[{\citenamefont{Weingartshofer
  et~al.}(1983)\citenamefont{Weingartshofer, Holmes, Sabbagh, and
  Chin}}]{WHS1983}
\bibinfo{author}{\bibfnamefont{A.}~\bibnamefont{Weingartshofer}},
  \bibinfo{author}{\bibfnamefont{J.~K.} \bibnamefont{Holmes}},
  \bibinfo{author}{\bibfnamefont{J.}~\bibnamefont{Sabbagh}}, \bibnamefont{and}
  \bibinfo{author}{\bibfnamefont{S.~L.} \bibnamefont{Chin}},
  \bibinfo{journal}{J.\ Phys.\ B} \textbf{\bibinfo{volume}{16}},
  \bibinfo{pages}{1805} (\bibinfo{year}{1983}).

\bibitem[{\citenamefont{Tizei et~al.}(2015)\citenamefont{Tizei, Lin, Mukai,
  Sawada, Lu, Li, Kimoto, and Suenaga}}]{TLM15}
\bibinfo{author}{\bibfnamefont{L.~H.~G.} \bibnamefont{Tizei}},
  \bibinfo{author}{\bibfnamefont{Y.-C.} \bibnamefont{Lin}},
  \bibinfo{author}{\bibfnamefont{M.}~\bibnamefont{Mukai}},
  \bibinfo{author}{\bibfnamefont{H.}~\bibnamefont{Sawada}},
  \bibinfo{author}{\bibfnamefont{A.-Y.} \bibnamefont{Lu}},
  \bibinfo{author}{\bibfnamefont{L.-J.} \bibnamefont{Li}},
  \bibinfo{author}{\bibfnamefont{K.}~\bibnamefont{Kimoto}}, \bibnamefont{and}
  \bibinfo{author}{\bibfnamefont{K.}~\bibnamefont{Suenaga}},
  \bibinfo{journal}{Phys.\ Rev.\ Lett.} \textbf{\bibinfo{volume}{114}},
  \bibinfo{pages}{107601} (\bibinfo{year}{2015}).

\bibitem[{\citenamefont{Asenjo-Garcia and {Garc\'{\i}a de
  Abajo}}(2013)}]{paper221}
\bibinfo{author}{\bibfnamefont{A.}~\bibnamefont{Asenjo-Garcia}}
  \bibnamefont{and} \bibinfo{author}{\bibfnamefont{F.~J.}
  \bibnamefont{{Garc\'{\i}a de Abajo}}}, \bibinfo{journal}{New\ J.\ Phys.}
  \textbf{\bibinfo{volume}{15}}, \bibinfo{pages}{103021}
  (\bibinfo{year}{2013}).

\bibitem[{\citenamefont{Faisal}(1987)}]{F1987}
\bibinfo{author}{\bibfnamefont{F.~H.~M.} \bibnamefont{Faisal}},
  \emph{\bibinfo{title}{Theory of Multiphoton Processes}}
  (\bibinfo{publisher}{Springer}, \bibinfo{address}{New York},
  \bibinfo{year}{1987}).

\bibitem[{Flo()}]{Floquetqtow}
\bibinfo{note}{Notice that a $-$ sign is introduced in
  $\hbar\omega=-\hbar(q-q_0)v$ to interpret a negative value of $q-q_0$ as a
  positive energy loss $\hbar\omega$ in the observed EELS spectrum of the ZLP.}

\end{thebibliography}

%\clearpage %--- optional
%\pagebreak \onecolumngrid \section*{SUPPLEMENTARY FIGURES} %---SI---arxiv optional

\end{document}